\documentclass[a4paper,11pt]{article}
\renewcommand{\baselinestretch}{2}
\usepackage{multirow}
\usepackage{fontawesome5} 
\usepackage{graphicx}

\usepackage{lscape}

\usepackage{diagbox}

\usepackage[T1]{fontenc}
\usepackage[latin1]{inputenc}
\usepackage{latexsym} 
\usepackage{amssymb}
\usepackage{fancyvrb}
\usepackage{algorithmic}
\usepackage{tikz}

\usepackage[T1]{fontenc}
\usepackage[latin1]{inputenc}
\usepackage[english]{babel}
\usepackage{latexsym} 
\usepackage{amssymb}
\usepackage{amsmath}
\usepackage{array}
\usepackage{bbding} 
\usepackage{mathpazo} 

\usepackage{amsthm}
\usepackage{xcolor}

\usepackage{apacite}

\usepackage{setspace}
\usepackage[font=small]{caption}
\usepackage{rotating}

\usepackage{subcaption}
\usepackage{tikz}
\usetikzlibrary{backgrounds}

\usetikzlibrary{shapes.geometric}
\usetikzlibrary{arrows,automata}

\usepackage{multirow}

\usepackage{color}
\usepackage{colortbl}

\usepackage{pgf}
\usepackage{mathrsfs}
\usetikzlibrary{arrows}

\usepackage{algorithmic}
\usepackage{longtable}
\usepackage{arydshln}
\usepackage{hyperref}
\usepackage{pdflscape}
\allowdisplaybreaks
\usepackage[margin=1 in]{geometry}
\usepackage{natbib}
\PassOptionsToPackage{numbers, compress}{natbib}
\usepackage{pdflscape}

\theoremstyle{plain}

\newtheorem{theorem}{Theorem}[section]
\newtheorem{proposition}[theorem]{Proposition}

\newtheorem{example}[theorem]{Example}

\theoremstyle{definition}

\theoremstyle{remark}

\usepackage{floatrow}
\newfloatcommand{capbtabbox}{table}[][\FBwidth]

\usepackage{blindtext}

\newcommand{\addQEDstyle}[2]{\AtBeginEnvironment{#1}{\pushQED{\qed}\renewcommand{\qedsymbol}{#2}}\AtEndEnvironment{#1}{\popQED}}
\addQEDstyle{example}{$\triangle$}

\usepackage{siunitx}

\newcolumntype{R}[2]{%
    >{\adjustbox{angle=#1,lap=\width-(#2)}\bgroup}%
    l%
    <{\egroup}%
}
\allowdisplaybreaks 
\begin{document}

\title{\textbf{A contemporary approach on revisited cost allocation using airport games: the effects of code-sharing}}

\author{Alejandro Saavedra-Nieves$^1$, M. Gloria Fiestras-Janeiro$^2$}

\date{\footnotesize \emph{
		$^1$ \underline{Corresponding author}. CITMAga. Departamento de Estat{\'i}stica, An{\'a}lise Matem{\'a}tica y Optimizaci{\'o}n, Universidade de Santiago de Compostela. \texttt{alejandro.saavedra.nieves@usc.es}\\
		$^2$   CITMAga. Departamento de Estat{\'i}stica e Investigaci{\'o}n Operativa, Universidade de Vigo. \texttt{fiestras@uvigo.gal} \\		}}

\maketitle

\onehalfspacing

\renewcommand{\baselinestretch}{1.4}

\abstract{\textcolor{black}{An important operational aspect in airport management is the allocation of fees to aircraft movements on a runway, whether operated by separate operators or under code-sharing agreements. This paper analyses the problem of determining  fees under code-sharing of the movements at an airport from a game theoretic perspective. In particular, we propose the configuration value for games with coalition configuration as the mechanism for allocating operating costs. We provide the exact expression of this game-theoretic solution for airport games, which depends only on the parameters of the associated airport problem. For this purpose, we consider a new and natural game-theoretic characterization of the configuration value. Finally, for the specific context of airport games, we apply it to a real case as a mechanism to determine the aircraft fees at a Spanish airport in a code-sharing scenario.}

\maketitle

\vspace{0.15cm}

\noindent {\bf Keywords:} Aircraft fees, code-sharing, coalition configuration, configuration value, axioms

\vspace{0.15cm}
\section{Introduction}

The aviation industry plays a pivotal role in global connectivity, with millions of passengers relying on efficient and cost-effective air travel. Airline alliances are a collaborative response to today's society's demand  for mobility  and operational efficiency. Such structures allow airlines to expand their network reach without significant investment, offering seamless travel through code-sharing and coordinated schedules. Alliances aim to increase  customer loyalty by improving convenience and route options, while enhancing the competitiveness of member airlines. However, challenges still remain, including the coordination of different operational systems and the potential dilution of brand identity. These disadvantages are offset to some extent by a reduction in operating costs through joint initiatives in areas such as maintenance, fuel purchasing and airport facilities. 

\textcolor{black}{In a collaborative scheme, considering a game-theoretic framework to model the cooperation of agents derives a realistic fitting when using transferable utility (TU) games. Focusing on transport infrastructures, this approach ensures that every potential alliance leads to an improvement in performance (e.g., in terms of costs or profits) compared to the marginal contributions of individual elements. For example, \cite{rosenthal2017cooperative} studied rapid transit networks; \cite{hadas2017approach}  also analyzed network connectivity in transport; \cite{algaba2019horizontal} analyzed the profit problem in an intermodal transport system; and \cite{gusev2020vertex} addressed the problem of distributing surveillance devices across a road network under this approach. However, cooperative game theory has shown considerable utility in identifying different methods of allocating fixed costs \citep{young1985cost}. We highlight its extensive use in the allocation of fixed costs, particularly, in transport contexts. For example, \cite{henriet1996traffic} studied cost allocation problems in transport networks. Similarly, \cite{fragnelli2000share}, \cite{norde2002balancedness} and \cite{costa2016polynomial} examined fixed cost allocation in railway infrastructure. The allocation of highway costs among users has also been studied from \cite{kuipers2013sharing}, \cite{gomez2024cost}, and \cite{wu2024highway}, and \cite{estan2021allocate} proposed an allocation method inspired by tram systems.}

The study of airport situations using TU games has also received significant attention. For instance, \cite{netessine2005revenue} and \cite{ccetiner2013assessing} focused on revenue sharing and management in the airline industry. However, this paper examines a fundamental class of cost-sharing problems, collectively termed  \emph{airport problems}. These situations are characterized by the existence of multiple airlines that use a runway of an airport and assuming that larger planes require longer runways. Once the runway is built to accommodate the largest plane, no additional cost is required for smaller planes. The key question is how to allocate the construction cost among all airlines. This hypothesis serves as a general framework for many other cost-sharing problems of this nature. We mention the reference \cite{thomson2024cost}, which can be seen in a recent survey, although \cite{baker1965airport} and \cite{thompson1971} are considered seminal papers on this topic. The game-theoretic approach associates with each airport problem an airport game, that is a TU game that assigns to each possible group of movements at the airport the costs of the runway needed at the airport. The application of solution concepts for TU games has been extensively studied in the literature to determine  fees for aircraft within the framework of airport games. \textcolor{black}{Table \ref{summary} lists the main solutions concepts for TU games already considered as a cost allocation mechanism in the context of airport games. In each case, the exact computation of the solution concept in the context of airport games avoids the use of its general definition, which in most cases involves computational complexity of exponential order.}

 \begin{table}[!h]
 	\begin{center}
 		\resizebox{0.9\textwidth}{!}{	
 		\begin{tabular}{|p{3.5cm}|p{4.5cm}|p{7cm}|}	\hline
Solution-concept & TU games& Airport games \vspace{0.15 cm}\\\hline\hline
 The \emph{Shapley value} & \cite{Shapley1953} & \cite{littlechild1973simple} \vspace{0.15 cm}\\
 The \emph{Banzhaf value} & \cite{banzhaf1964weighted} & \cite{saavedra2019contributions}\vspace{0.15 cm}\\
 The \emph{nucleolus} & \cite{schmeidler1969nucleolus} & \cite{littlechild1974simple}\vspace{0.15 cm}\\
 The $\tau$-\emph{value} & \cite{tijs1981bounds} & \cite{tijs1986game,driessen1988cooperative}\vspace{0.15 cm}\\
 The \emph{core-center} & \cite{gonzalez2007natural} &\cite{gonzalez2016airport}\\\hline
 \end{tabular}}
\caption{Summary of references on solution concepts in airport games.} \label{summary}
\end{center}
 \end{table}

Drawing from the given references, research on airport games has branched out extensively in multiple directions over recent decades. Examples of alternative references are \cite{dubey1982shapley} and \cite{hou2018note}, which explore scenarios that differ slightly from the original airport games. In these cases, the relationship between airport fees and the Shapley value of the resulting games is analyzed. While the existing airport game models in the literature are very valuable from a theoretical point of view, in practice it was noticeable that they have overlooked  a crucial aspect of determining aircraft fees, namely the organization of aircraft within airlines. Rather than treating airplanes as isolated units, they should be viewed as part of an airline. As discussed above, larger airlines, in particular, would have greater leverage to negotiate cost advantages, such as a discount, than smaller ones. To model such situations, TU games with a priori unions have been used, which allow the organization of aircraft into airlines to be taken into account. In this context, the use of  solution concepts for the resulting games with a priori unions specifies new systems of fees for aircraft, now taking into account their structure into airlines. In particular, we mention the case of the \emph{Owen value} for TU games with a priori unions \citep{owen1977values}, which formally extends the Shapley value \citep{Shapley1953} to this new context, and which was extensively studied for airport games in \cite{vazquez1997owen}. In line with this reference, \cite{saavedra2019contributions} similarly provide the exact expression for the Banzhaf-Owen value in airport games. The \emph{Banzhaf-Owen value} \citep{owen1982modification} is an extension of the Banzhaf value to situations where there exists a partition structure over  the set of players. \cite{casas2003extension} also extend the $\tau$-value to games with a priori unions, focusing on the case of airport games.

However, aircraft fees resulting from taking into account airlines structure fail to reflect the real collaborative dynamics of modern aviation. In such scenarios, airlines with small number of  operations may be reluctant to enter into partnerships because they do not bear the operating costs of larger operations. Therefore, a more nuanced approach, such as incorporating code-sharing where multiple airlines benefit from a single operation, allows operational costs to be shared in a more comprehensive and equitable framework that reflects the collaborative nature of modern aviation. \textcolor{black}{This is particularly relevant in the context of global airline alliances like Star Alliance, SkyTeam, or Oneworld, where member airlines routinely share flights, networks, and services. By aligning aircraft charges with the actual beneficiaries of code-shared flights, the system ensures that only those airlines involved contribute proportionately and fairly, thereby encouraging further cooperation and maximizing overall airport efficiency and revenue. These dynamics raise a critical question for airport  schemes: how should maintaining fees be fairly allocated among code-sharing airlines? Addressing this issue is essential to ensuring that airport cost-recovery mechanisms remain consistent with the collaborative and networked nature of contemporary airline operations.} 

In this paper, we analyze a new approach to the problem of determining aircraft fees under code-sharing using cooperative game theory. To the best of our knowledge, such a problem has not been considered in the literature under this perspective. Specifically, we will consider the original cooperative framework of airport games introduced in \cite{littlechild1973simple}. The main  novelty of our proposal lies in the fact  that each movement of a plane can be operated by more than one airline and thus, such airline would pay the associated fee for the movement. It allows the use of the model of  \emph{games with coalition configuration} \citep{albizuri2006configuration} to represent these situations. As usual in cooperative game theory, the aim is to distribute the global costs among the set of players involved in a reasonable way. We innovatively posit to use the configuration value \citep{albizuri2006configuration}, as {an alternative tool to the Shapley value} or the Owen value for games with a priori unions, that \cite{littlechild1973simple} and \cite{vazquez1997owen}, respectively, considered to define new aircraft fees for airport games. In particular, this approach  allows us to take into account the organization of flights under code-sharing agreements \textcolor{black}{for airport cost operation allocation and that, grounded in well-established axioms, make the configuration value well-suited for application in this new context.} As a main contribution,  despite the high computational complexity involved in its exact computation, we obtain its exact expression for the specific case of airport games with coalition configuration,  that can be directly obtained from the parameters of the associated airport problem. Previously, we also provide a new characterization of the configuration value for general TU games with coalition configuration, which we relied on to obtain its explicit expression in the specific context of airport games. 
Finally, as an application, we determine the fees for aircraft movements at the airport of Santiago de Compostela (Galicia, Spain) for two non-working days, comparing the results based on the configuration value with those based on the Owen value and the Shapley value studied in the literature.

This paper is structured as follows. Section \ref{sec:airportgames} introduces the key concepts from cooperative game theory that are essential for addressing the problem under consideration. In Section \ref{sec:conf_value}, we revisit the configuration value for games with coalition configuration and we offer a novel axiomatic characterization for it, which is particularly appealing in the context of determining aircraft fees. In Section \ref{exact}, we present the exact expression for the configuration value in airport games taking into account the presence of coalition configurations in the set of movements. Our proposed cost allocation method is then applied to a real-world airport scenario in Section \ref{sec:results}. Finally, Section \ref{sec:conclusions} provides some concluding remarks.

\section{Airport games}\label{sec:airportgames}


The problem of allocating the costs of building and using an airport movement area from a game-theoretic perspective has been studied extensively in the literature. As discussed in \cite{littlechild1977aircraft}, in an airport context, a movement is a take-off or a landing, that involves the runways, taxiways, and apron, distinct from the terminal area. Since the cost of building or maintaining a runway depends largely on the "largest" aircraft it is designed to handle, such costs can be structured in an interesting way. {Notice that the length of the runway required for the operation of each aircraft depends on its characteristics and, particularly, on its take-off weight.} Thus, the cost of subsequent use of the runway is proportional to the weight of all  movements of aircraft. Consequently, while the allocation of variable costs for each aircraft arriving at or departing from the airport is relatively straightforward, the challenge lies in distributing the fixed costs associated with aircraft movements.

From a mathematical perspective, an \emph{airport problem} is defined as a cost allocation situation where $N$ denotes the set of movements of the involved airplanes.  Formally, we assume the existence of several types of planes, ${\cal T}=\{1,\dots,|{\cal T}|\}$, which operate in an airport for a fixed period of time. For each type of movement $\tau\in {\cal T}$, there exists a cost $c_{\tau}$ satisfying that $c_0=0\leq c_1\leq \ldots \leq c_{|\cal T|}$, and $N_{\tau}$ denotes the set of movements of planes of type $\tau \in {\cal T}$, with $N=\cup_{\tau\in {\cal T}}N_{\tau}$.

A cooperative game theoretic perspective can be considered for the allocation of the fixed costs in an airport problem. A {cost game} is given by a pair $(N, c)$, where $N$ denotes the set of players and $c$ is the {characteristic function} $c:2^N\longrightarrow \mathbb{R}$ satisfying $c(\emptyset) = 0$. For each coalition $S\subseteq N$, $c(S)$ represents the cost of the cooperation of the players in $S$. In this context, we  use the cost game $(N,c)$, associated to any airport problem, that is defined in \cite{littlechild1973simple} by identifying the players as the individual movements of airplanes.  Formally, an \emph{airport game} is given by $(N,c)$ where $N$ denotes the set of movements and $c$ is a map that assigns, for each coalition $S\subseteq N$,\begin{equation}\label{airportgame}
    c(S)=\max\{c_{\tau}: S\cap N_{\tau}\neq \emptyset\}.
\end{equation}
Such cost corresponds to the ``largest'' \mbox{movement} belonging to coalition $S$. Recall that the largeness of a movement refers to the runway length required for its landing/take-off. Since the take-off weight of each aircraft model usually determines this length, an aircraft will be longer when its take-off weight is greater.

The use of game theoretic allocation rules was already considered in literature for covering the task of allocating the costs in airport problems. We focus on the case of the Shapley value \citep{Shapley1953} for $(N,c)$ and their extensions. 
The next formula provides the expression of the Shapley value for an airport game $(N,c)$ in terms of the parameters of the model. Let $(N,c)$ be an airport game and take $i\in N$ such that it corresponds to a movement of type $\tau(i)\in \{1,\dots,|{\cal T}|\}$. The Shapley value of the airport game $(N,c)$ (cf. \citealp{littlechild1973simple}) assigns to each movement $i\in N${\color{black}\begin{equation}\label{shapley}
	Sh_{i}(N,c)=\sum_{t=1}^{\tau(i)}\frac{c_{t}-c_{t-1}}{ {|N_{\geq t}|}}
	\end{equation}}
such $c_0=0$, $\tau(i)$ denotes the type of the airplane of movement $i$, and $N_{\geq t}=\cup_{\tau=t}^{|\cal T|}N_{\tau}$ is the set of movements of planes of type $t$ or larger. The underlying idea was originally described in \cite{baker1965airport} and \cite{thompson1971}. First, the runway costs attributed to the smallest aircraft type are equally divided  among the total number of movements for all aircraft. This initial part of the allocation scheme appears as fair, as all movements utilize this portion of the runway. Next, the incremental cost of the runway for the second smallest type of aircraft is divided equally over the number of movements for all aircraft types except the smallest. This process continues until the incremental cost of the largest aircraft type is equally divided  among the number of movements of the largest type of aircraft. 

Such approach is insightful but overlooked that aircraft movements were typically tied to airline agreements rather than individual flights. \cite{vazquez1997owen} also analyzed the cost allocation problem arising from the grouping of the aircraft movements into airlines. For this purpose, cost games with a priori unions are considered. A cost game with a priori unions is a triplet $(N,c,P)$ where $(N,c)$ is a cost game and $P=\{P_1,\dots,P_m\}$ is a partition of $N$ describing a structure of cooperation for the players. In airport settings, such structures model the organization of movements in airlines. Thus, if $\mathcal{A}=\{1,\dots,A\}$ represents the set of airlines operating at the airport, $P=\{P_1,\dots,P_A\}$ denotes a partition over the set of movements $N$, with $P_a$ being the set of movements operated by airline $a\in \mathcal{A}$.

Take $(N,c,P)$ a cost game with a priori unions. The Owen value for  $(N,c,P)$ (cf. \citealp{owen1977values}) can be seen as an extension of the Shapley value under the presence of a system of unions.  In what follows, we denote by $P^{i}$ the union in $P$ to which $i$ belongs. Then, the Owen value assigns, for every $i\in N$ and for every $(N,c,P)$, 

\begin{equation}\label{owvalue}\small
Ow_i(N,c,P)=\sum_{\mathcal{C}\subseteq P\setminus P^{i}}\sum_{\substack{S\subseteq P^i\\i\in S}}\frac{|\mathcal{C}|!(m-|\mathcal{C}|-1)!}{m!}\frac{(|S|-1)!(|P^i|-|S|)!}{|P^i|!}\big(c(A_{\mathcal{C}}\cup S)-c(A_{\mathcal{C}}\cup (S\setminus \{i\})\big),\end{equation}
where $A_{\mathcal{C}}=\underset{P_r\in \mathcal{C}}{\cup}P_r$. 

In airport settings, \cite{vazquez1997owen} specifically provided a simple expression for the Owen value of an airport game with a priori unions $(N,c,P)$. Take $i\in N$, that corresponds to a movement of a plane of type $\tau(i)\in \{1,\dots,{\cal T}\}$ operated by a unique airline $a\in \{1,\dots,A\}$. Then, the Owen value of  movement $i$ is
{\color{black}\begin{equation}\label{owen}
	Ow_{i}(N,c,P)=\sum_{t=1}^{\tau(i)}\frac{c_{t}-c_{t-1}}{{|\mathcal{A}_{\geq t}|}\ {|N^{a}_{\geq t}|}}\end{equation}}
being $c_0=0$, $N_{\geq t}^{a}=\cup_{\tau=t}^{|\cal T|}N_{\tau}\cap P_{a}$ the set of movements of planes of type $t$ or larger operated by the airline $a$ and $\mathcal{A}_{\geq t}=\{a\in \{1,\dots,A\}: N_{\geq t}^a\neq \emptyset\}$ is the set of airlines with movements of type $t$ or larger. 
The allocation process divides the costs of constructing each segment of the runway based on the types of movements each airline operates. The cost for constructing the initial section of the runway, given by $c_1$,  which is incurred by all types of airplanes, is allocated equally among all airlines, and within each airline, this cost is further allocated equally among its movements. For each subsequent runway segment, this amount (e.g., $c_2-c_1$ for the second part) is equally allocated among the airlines that operate airplanes requiring that segment, and similarly, this amount is equally divided among the applicable movements within each airline. This process continues until the total cost $c_{|\mathcal{T}|}$ is allocated among all airplane movements at the airport. 

\section{The configuration value}\label{sec:conf_value}

In multi-agent systems, agents often prefer to cooperate with certain others due to shared interests, making coalition structures a fundamental tool for analysis. However, such framework may inadequately represent some bargaining scenarios where agents do not necessarily organize themselves into mutually exclusive coalitions. As \cite{aumann2003endogenous} noted, relationships between agents are not always transitive. To address such complexities, the notion of a coalition configuration is needed. A coalition configuration is formally defined as a family of coalitions that are not necessarily disjoint, the union of which is the grand coalition. In other words, agents are allowed to form coalitions that are not mutually exclusive in order to achieve a better bargaining position. 


{The following subsections present a game-theoretic analysis of the configuration value. In Subsection~\ref{subsec31}, we first provide its definition and the  characterization given in \cite{albizuri2006configuration}. Subsequently, Subsection~\ref{sec:characterization} introduces a new characterization based on its relationship with the Owen value for a game with a priori unions.}

\subsection{The game-theoretic framework}\label{subsec31}

Let $N$ be the set of players. A \emph{coalition configuration} of $N$ is a family $\mathcal{B}=\{B_1,\cdots,B_m\}$ of coalitions of $N$ satisfying that $\overset{m}{\underset{r=1}{\cup}}B_r=N$. For every $i\in N$, we denote by $\mathcal{B}^i=\{B_p\in\mathcal{B}\ : \ i\in B_p\}$ the set of those elements of $\mathcal{B}$ to which player $i$ belongs to. For any cost game $(N,c)$ and a coalition configuration $\mathcal{B}$, we denote by $(N,c,\mathcal{B})$ the associated game with coalition configuration. 

A relevant issue, as usual in cooperative game theory, deals with coalitions and allocations, considering groups of agents willing to allocate
the joint benefits derived from their cooperation. Let $(N,c,\mathcal{B})$ be a  game with coalition configuration. An allocation for $(N,c,\mathcal{B})$ 
is an element $x\in \mathbb{R}^{|N|}$. An allocation rule for games with coalition configuration is a map $\psi$, which assigns to every game with coalition configuration $(N,c,\mathcal{B})$ an allocation $\psi(N,c,\mathcal{B})\in \mathbb{R}^{|N|}$.

Take $(N,c,\mathcal{B})$ a cost game with coalition configuration. The \emph{configuration value} (cf. \citealp{albizuri2006configuration}) is defined as the allocation rule  that assigns, for every $i\in N$ and every $(N,c,\mathcal{B})$,
\begin{equation}\label{confvalue}\small
CV_i(N,c,\mathcal{B})=\sum_ {B_q\in\mathcal{B}^i}\sum_ {\substack{\mathcal{C}\subseteq \mathcal{B}\\\mathcal{C}\cap\mathcal{B}^i=\emptyset}}\sum_{\substack{S\subseteq B_q\\i\in S}}\frac{|\mathcal{C}|!(m-|\mathcal{C}|-1)!}{m!}\frac{(|S|-1)!(|B_q|-|S|)!}{|B_q|!}\big(c(A_{\mathcal{C}}\cup S)-c(A_{\mathcal{C}}\cup (S\setminus \{i\})\big),\end{equation}
where $A_{\mathcal{C}}=\underset{B_r\in \mathcal{C}}{\cup}B_r$ for every $\mathcal{C}\subseteq \mathcal{B}$. Note that the configuration value coincides with the Owen value of the corresponding game with a priori unions in the special case of $\mathcal{B}$ being a partition.  So, it also corresponds to the Shapley value when $\mathcal{B}=\{\{i\} : \ i\in N\}$.

Below, we list some interesting properties, extracted from \cite{albizuri2006configuration}, to be satisfied for any allocation rule $\psi$ defined for any cost game with coalition configuration $(N,c,\mathcal{B})$. 
\begin{itemize}
    \item[\textbf{(EFF)}] \textbf{Efficiency.} An allocation rule $\psi$ satisfies \emph{efficiency} if, for any $(N,c,\mathcal{B})$, it holds that $\underset{i\in N}{\sum}{\psi_i(N,c,\mathcal{B})}=c(N)$.
    \item[\textbf{(NPP)}] \textbf{Null player property.} An allocation rule $\psi$ satisfies \emph{null player property} if, for any $(N,c,\mathcal{B})$, it holds that $\psi_i(N,c,\mathcal{B})=0$ if $i$ is a null player. Recall that $i$ is said to be a null player in $c$ if $c(S\cup\{i\})-c(S)=0$ for all $S\subseteq N$.
     \item[\textbf{(L)}] \textbf{Linearity.} An allocation rule $\psi$ satisfies \emph{linearity} if, for any $(N,c,\mathcal{B})$ and $(N,c^{\prime},\mathcal{B})$ and $\lambda,\mu\in\mathbb{R}$, it holds that $\psi(N,\lambda c+\mu c^{\prime},\mathcal{B})=\lambda\psi(N,c,\mathcal{B})+\mu \psi(N,c^{\prime},\mathcal{B}).$
 \item[\textbf{(A)}] \textbf{Anonymity.} If $\sigma$ is a permutation of $N$ such that $\sigma(B_p)=B_p$ for every $B_p\in \mathcal{B}$ then, for every $i\in N$, it holds $\psi_i(N,\sigma c,\mathcal{B})=\psi_{\sigma(i)}(N,c,\mathcal{B})$, where $(N,\sigma c)$ is the TU game defined by $(\sigma c)(S)=c(\sigma(S))$ for every $S\subseteq N$.
  \item[\textbf{(CS)}] \textbf{Coalitional symmetry.} An allocation rule  $\psi$ satisfies \emph{coalitional symmetry} if, for any partition $\mathcal{B}$ of $N$ and for any pair $B_p,B_q\in\mathcal{B}$ such that for every $\mathcal{C}\subseteq \mathcal{B}\setminus\{B_p,B_q\}$ satisfying $c(B_p\cup \underset{B_r\in \mathcal{C}}{\cup}B_r)=c(B_q\cup \underset{B_r\in \mathcal{C}}{\cup}B_r)$, it holds that $\underset{i\in B_p}{\sum}{\psi_i(N,c,\mathcal{B})}=\underset{i\in B_q}{\sum}{\psi_i(N,c,\mathcal{B})}$.
\end{itemize}
Before introducing the last axiom, we introduce some notation taken from \cite{albizuri2006configuration}. Given a pair of players $i,j\in N$, we define the game $(N\setminus\{j\},c_{i\setminus j})$ by $$c_{i\setminus j}(S)=\begin{cases}
	c(S), & \mbox{if } i\notin S; \\
	c(S\cup\{j\}), & \mbox{if } i\in S.
\end{cases}$$
We also denote by $\mathcal{B}_{i\setminus j}$ the coalition configuration on $N\setminus\{j\}$ specified by $\mathcal{B}_{i\setminus j}=(\mathcal{B}\setminus \mathcal{B}^j)\cup\{(B_r\setminus\{j\})\cup\{i\} : \ B_r\in \mathcal{B}^j\}$. Two players $i,j\in N$ are doubles in $(N,c,\mathcal{B})$ if (i) $c(S\cup \{i\})=c(S\cup \{j\})$ for every $S\subseteq N$ and (ii) $S\cup\{i\}\in \mathcal{B}^i$ implies that $S\cup \{j\}, S\cup\{i,j\}\notin \mathcal{B}^j$, for every $S\subseteq N\setminus\{i,j\}$.
\begin{itemize}
    \item[\textbf{(M)}] \textbf{Merger.} If $i,j\in N$ are doubles in $(N,c,\mathcal{B})$ and $\mathcal{B}^i\cap\mathcal{B}^j=\emptyset$, then for every $k\in N\setminus \{i,j\}$, $\psi_k(N,c,\mathcal{B})=\psi_k(N,c_{i\setminus j},\mathcal{B}_{i\setminus j})$.
\end{itemize}
\textcolor{black}{These properties justify using the configuration value as a cost allocation mechanism in multi-agent problems under a coalition configuration. They ensure that the resulting allocation  adapts fairly and consistently to the structural features of the problem.  Efficiency (EFF) guarantees that the total cost is fully distributed while the null player property (NPP) ensures that no cost is assigned to non-contributing participants. Linearity (L) enables the rule to behave predictably under linear combinations of cost functions and anonymity (A) ensures that allocations depend solely on the cost function and coalition structure rather than on the role or position of players. Coalitional symmetry (CS) enforces equal treatment for coalitions with equivalent contributions, and the merger property (M) imposes equal treatment between coalitions with the same cost contribution capabilities. Together, these properties ensure that the allocation mechanism is equitable and robust under changes in the coalition configuration.}

\cite{albizuri2006configuration}  characterized the configuration value for games with coalition configuration as follows.

\begin{theorem}\label{thalb}(\citealp{albizuri2006configuration}) Let $(N,c,\mathcal{B})$ a game with coalition configuration. An allocation rule $\psi$ satisfies (EFF), (NPP), (L),
(A), (CS) and (M) if and only if it is the configuration value $CV(N,c,\mathcal{B})$.    
\end{theorem}

\subsection{A new characterization of the configuration value}\label{sec:characterization}

In this section we provide a new characterization of the configuration value that consider its relation with the Owen value for a game with a priori unions. Specifically, \cite{albizuri2006configuration} proved that Merger property (M) also implies that the configuration value can be obtained by means of  the Owen value for an associated TU game with a priori unions.   

To formalize this idea, we introduce below some required notation. Take $(N,c,\mathcal{B})$ a game with coalition configuration. For every fixed $i\in N$, with $i\in B_q$, $i(q)$ denotes the representative of $i$ in $B_q$ in such a way that $i(q)\neq j(r)$ if $i\neq j$ or $q\neq r$. Denote $M_i=\{i(q)\ : \ B_q\in \mathcal{B}^i\}$, do $\bar{N}=\cup_{i\in N}M_i$ and $\bar{B}_q=\{i(q) : i\in B_q\}$ for all $q\in \{1,2,\dots,m\}$. Thus, we obtain by construction the partition of $\bar{N}$\begin{equation}\label{Bbar}
    \bar{\mathcal{B}}=\{\bar{{B}}_1,\dots, \bar{{B}}_m\}.
\end{equation}
Besides, we define the game $(\bar{N},\bar{c})$ by
\begin{equation}\label{cbar}
    \bar{c}(\bar{S})=c(\{i\in N : M_i\cap \bar{S}\neq \emptyset\})\end{equation}
for every coalition $\bar{S}\subseteq \bar{N}$.

Thus, the configuration value for $(N,c,\mathcal{B})$ can be rewritten as
\begin{equation}\label{CV_OW}
CV_i(N,c,\mathcal{B})=\underset{q : B_q\in \mathcal{B}^i}{\sum}CV_{i(q)}(\bar{N},\bar{c},\bar{\mathcal{B}})=\underset{q : B_q\in \mathcal{B}^i}{\sum}Ow_{i(q)}(\bar{N},\bar{c},\bar{\mathcal{B}}).
\end{equation}

{The existing characterization of the configuration value \citep{albizuri2006configuration} uses axioms only involving the characteristic function of the associated TU games. In the case of the application of the configuration value in multi-agent cost situations, e.g. for determining fees  using airport games in the sense of \citep{vazquez1997owen}, it is more appealing to have a characterization of the configuration value  based on changes on the coalition configuration.}

From its definition, it is easy to see that the configuration value is a generalization of the Owen value, which, in turn, is interpreted as a generalization of the Shapley value. In this sense, TU games form a subclass of games with a priori unions when we consider a trivial system of unions, which, in turn, corresponds to a subclass of games with coalition configuration. Thus, the configuration value can be understood as a generalization of the Shapley value for situations in which the coalition configuration is either non-trivial or not a partition. To formalize this idea, we introduce the concept of configurational Shapley value.

{Below, we particularly focus on the case of configurational Shapley values.} A \emph{configurational Shapley value} is an allocation rule $\psi$ which assigns to any game with coalition configuration $(N,c,\mathcal{B})$ an allocation vector $\psi(N,c,\mathcal{B})\in\mathbb{R}^{N}$, such that for all games with a trivial coalition configuration (i.e. a configuration of $N$ whose elements are the individual players), it corresponds to the Shapley value of the associated game.

 To characterize the configuration value, we inspire in those natural properties for the Shapley value considered, among others, in \cite{Shapley1953} and \cite{dubey1982shapley}. Thus, we present a list of desirable properties to be satisfied for any allocation rule $\psi$ defined for any cost game with coalition configuration $(N,c,\mathcal{B})$  and we will show that the configuration value is the unique configurational Shapley value that satisfies these properties.

The first property refers to the consideration of each player in $N$ through its representatives in each element $B_q\in \mathcal{B}^i$.  Let  $(N,c,\mathcal{B})$ be a game with coalition configuration and take $\psi$ an allocation rule for $(N,c,\mathcal{B})$. Below, we present a first property on $\psi$.

\begin{itemize}
    \item[\textbf{(ADD-R)}] \textbf{Additivity in representatives.} An allocation rule $\psi$ satisfies \emph{additivity in representatives} if, for every $i\in N$, it holds that 
    \begin{equation}\label{ADDI}
\psi_i(N,c,\mathcal{B})=\underset{q : B_q\in \mathcal{B}^i}{\sum}\psi_{i(q)}(\bar{N},\bar{c},\bar{\mathcal{B}}),
\end{equation}
where $(\bar{N},\bar{c},\bar{\mathcal{B}})$ is the  game with a priori unions associated to $(N,c,\mathcal{B})$.
\end{itemize}

Next, we extend the property of  balanced contribution for the Owen value \citep{vazquez1997owen} to the case of the configuration value. Take $(N,c,\mathcal{B})$ and $(\bar{N},\bar{c},\bar{\mathcal{B}})$ the associated game with a priori unions. Fix player $k\in N$ such that $k\in B_p\in \mathcal{B}$. {We denote by $\bar{\mathcal{B}}_{-k(p)}$ the partition of $\bar{N}$ resulting from isolating the representative of player $k$ in $B_p$ in a single coalition formed by such player, after its removing from $\bar{B}_p$, i.e.,
\begin{equation}\label{Bmenoskp}
\bar{\mathcal{B}}_{-k(p)}=\big\{\{\bar{\mathcal{B}}\setminus \bar{B}_p\}\cup\{\bar{B}_p\setminus\{k(p)\}\cup\{k(p)\}\big\}.
\end{equation}}

\begin{itemize}
    \item[\textbf{(BC-CC)}]  \textbf{Balanced contributions for coalition configurations.} An allocation rule $\psi$ satisfies \emph{balanced contributions for coalition configurations} if, for any pair of players $i,j\in N$ such that $\mathcal{B}^i=\mathcal{B}^j$ it holds that
     \begin{equation}\label{BCCCprop}
        \psi_i(N,c,\mathcal{B})-\psi_j(N,c,\mathcal{B})=\underset{q : B_q\in \mathcal{B}^i}{\sum}\bigg(\psi_{i(q)}(\bar{N},\bar{c},\bar{\mathcal{B}}_{-j(q)})-\psi_{j(q)}(\bar{N},\bar{c},\bar{\mathcal{B}}_{-i(q)})\bigg),
    \end{equation}
    where $(\bar{N},\bar{c},\bar{\mathcal{B}}_{-k(p)})$ denotes the game with a priori unions  for every player $k\in B_p$ and for every   $B_p\in \mathcal{B}$.
\end{itemize}

{The property  states that if two players $i$ and $j$ belong to the same collection of coalitions in a given coalition configuration $\mathcal{B}$, then the difference between the allocations received by $i$ and $j$ corresponds to the sum, for all $B_p\in \mathcal{B}^i$, of the differences between the allocations of the representatives $i(p)$ and $j(p)$ in $(\bar{N},\bar{c},\bar{\mathcal{B}})$ when $j(p)$ and $i(p)$ respectively leave $B_p$ and are isolated themselves. As mentioned, this property extends the property of balanced contributions considered in \cite{vazquez1997owen} for the case of the existence of an a priori union system.}

Using this idea, we prove that the configuration value for games with coalition configuration satisfy the property of balanced contributions for coalition configurations.
\begin{proposition}\label{propBC}
    Let $(N,c,\mathcal{B})$ be a game with coalition configuration. Thus, the configuration value satisfies (BC-CC) property.
\end{proposition}
\begin{proof}
Take $(N,c,\mathcal{B})$ a game with coalition configuration and a pair of players $i,j\in N$ such that $\mathcal{B}^i=\mathcal{B}^j$. Thus, it holds that
\begin{equation*} \label{eq1}
\begin{split}
CV_i(N,c,\mathcal{B})-CV_j(N,c,\mathcal{B}) & = \underset{q : B_q\in \mathcal{B}^i}{\sum}Ow_{i(q)}(\bar{N},\bar{c},\bar{\mathcal{B}})-\underset{q : B_q\in \mathcal{B}^j}{\sum}Ow_{j(q)}(\bar{N},\bar{c},\bar{\mathcal{B}}) \allowdisplaybreaks\\
 & = \underset{q : B_q\in \mathcal{B}^i}{\sum}\bigg(Ow_{i(q)}(\bar{N},\bar{c},\bar{\mathcal{B}})-Ow_{j(q)}(\bar{N},\bar{c},\bar{\mathcal{B} })\bigg)\allowdisplaybreaks\\
 & = \underset{q : B_q\in \mathcal{B}^j}{\sum}\bigg(Ow_{i(q)}(\bar{N},\bar{c},\bar{\mathcal{B}}_{-j(q)})-Ow_{j(q)}(\bar{N},\bar{c},\bar{\mathcal{B}}_{-i(q)})\bigg)\allowdisplaybreaks\\
  & = \underset{q : B_q\in \mathcal{B}^j}{\sum}\bigg(CV_{i(q)}(\bar{N},\bar{c},\bar{\mathcal{B}}_{-j(q)})-CV_{j(q)}(\bar{N},\bar{c},\bar{\mathcal{B}}_{-i(q)})\bigg),
\end{split}
\end{equation*}
{where $\bar{\mathcal{B}}_{-k(p)}$, for every $k\in \bar{B}_q\in \bar{\mathcal{B}}$, is the partition in (\ref{Bmenoskp}) associated to $\mathcal{B}$. The first and the last equality are satisfied from the relation of the configuration value and the Owen value specified in (\ref{CV_OW}) and the third one, from the property of balanced contributions for the Owen value \citep{vazquez1997owen}, concluding the proof.}
\end{proof}

Finally, we extend the property of quotient game to the case of games with coalition configuration. For this purpose we consider the original notation in \cite{albizuri2006configuration}. Take $(N,c,\mathcal{B})$ a game with coalition configuration.  The game among coalitions on $\mathcal{B}$, denoted by $(M,c^{\mathcal{B}})$ with $M=\{1,\dots,m\}$, is defined for every subset of coalitions  $W\subseteq M$ by 
\begin{equation}\label{coalitionalgame}
    c^{\mathcal{B}}(W)=c(\cup_{k\in W}B_k).
\end{equation}

Next we formally introduce an interesting condition to be satisfied by each  allocation rule $\psi$ for games with coalition configuration.
\begin{itemize}
    \item[\textbf{(CQ)}] \textbf{Coalitional-quotient game property.} An allocation rule $\psi$ satisfies  \emph{coalitional-quotient game property} if for every game with coalition configuration $(N,c,\mathcal{B})$ and for every $B_q\in \mathcal{B}$, it holds that
    \begin{equation}
        \underset{j\in B_q}{\sum}\psi_{j(q)}(\bar{N},\bar{c},\bar{\mathcal{B}})=\psi_{B_q}(M,c^{\mathcal{B}},\hat{\mathcal{B}}),
    \end{equation}
    where $M=\{1,\dots,m\}$, $(M,c^{\mathcal{B}})$ is the game induced by the coalitions given in (\ref{coalitionalgame}), and $\hat{\mathcal{B}}$ is the trivial coalition configuration for the set of players $M$, i.e. $\hat{\mathcal{B}}=\{\{1\},\dots,\{m\}\}$.
\end{itemize}

{This property would ensure a kind of consistency of the allocation rule for games with coalition configuration in the sense that the sum of the individual allocations for the representatives in every $B_q\in \mathcal{B}$ when using $(\bar{N},\bar{c},\bar{\mathcal{B}})$ is equal to the total allocation for $B_q$ in the game among coalitions $(M,c^{\mathcal{B}},\hat{\mathcal{B}})$.}

The configuration value for games with coalition configuration satisfies (CQ) property. 

\begin{proposition}\label{propCQ}
    Let $(N,c,\mathcal{B})$ be a game with coalition configuration. The configuration value for $(N,c,\mathcal{B})$, $CV(N,c,\mathcal{B})$, satisfies (CQ) property.
\end{proposition}
\begin{proof} Take $(N,c,\mathcal{B})$ and consider the associated game with a priori unions $(\bar{N}, \bar{c}, \bar{\mathcal{B}})$,  with $(\bar{N}, \bar{c})$ being the game  in (\ref{cbar})  and $\bar{\mathcal{B}}$  the partition in (\ref{Bbar}). Clearly, the configuration value for $(\bar{N}, \bar{c}, \bar{\mathcal{B}})$ coincides with the Owen value. Since the Owen value satisfies the property  of quotient game \citep{vazquez1997owen}, we have
\begin{equation*}
       \underset{j\in B_q}{\sum}Ow_{j(q)}(\bar{N},\bar{c},\bar{\mathcal{B}})=Ow_{B_q}(M,c^{\mathcal{B}},\hat{\mathcal{B}}),
\end{equation*}
where $\hat{\mathcal{B}}$ is the trivial coalition configuration for the set of players $M$ whose elements are the individual players. Moreover, $Ow_{j(q)}(\bar{N},\bar{c},\bar{\mathcal{B}})=CV_{j(q)}(\bar{N},\bar{c},\bar{\mathcal{B}})$ and then, $\underset{j\in B_q}{\sum}CV_{j(q)}(\bar{N},\bar{c},\bar{\mathcal{B}})=CV_{B_q}(M,c^{\mathcal{B}},\hat{\mathcal{B}})$.
\end{proof}

The following theorem states that the  additivity in representatives, the balanced contributions and the coalitional-quotient game properties characterize a unique configurational Shapley value that is the configuration value.  \textcolor{black}{It is worth noting that this result provides an alternative characterization to the one in  \cite{albizuri2006configuration} which extends the characterization of the Owen value proposed in \cite{vazquez1997owen} in the context of the existence of a priori unions.}

\begin{theorem}\label{ThCV}
    The configuration value is the unique configurational Shapley value satisfying additivity in representatives (ADD-R), balanced contributions for coalition configurations (BC-CC) and coalitional-quotient game (CQ) properties.
\end{theorem}
\begin{proof}
We prove the (a) uniqueness and (b) existence of an allocation rule $\psi$ for a game with coalition configuration $(N, c, \mathcal{B})$ that is a configurational Shapley value and that it satisfies (ADD-R), (BC-CC), and (CQ) properties.

If $\mathcal{B}$ is a coalition configuration for $N$ such that
$\underset{\substack{B_p,B_q\in \mathcal{B}\ : \ B_p\neq B_q}}{\sum}|B_p\cap B_q|=0$, then $\mathcal{B}$ reduces to a partition of $N$.  \cite{vazquez1997owen} proved the existence and uniqueness of a coalitional Shapley value satisfying  (BC-CC) and (CQ) properties for the case of games with a priori unions, which is the Owen value.

Now, we focus on the uniqueness and the existence of an allocation rule for games with coalition configuration in the more general case, where there exists a coalition configuration $\mathcal{B}$ for $N$ such that
$\underset{\substack{B_p,B_q\in \mathcal{B}\ : \ B_p\neq B_q}}{\sum}|B_p\cap B_q|>0$.

    \begin{itemize}
        \item[(a)] \underline{Uniqueness}. First, we assume the existence of two different configurational Shapley values, $\psi^1$ and $\psi^2$, satisfying (ADD-R), (BC-CC) and (CQ) properties. That is, there exist a game with coalition configuration $(N,c,\mathcal{B})$ and a player $i\in N$ such that $\psi^1_i(N,c,\mathcal{B})\neq \psi^2_i(N,c,\mathcal{B})$. 

        Since $\psi^1$ and $\psi^2$  satisfy (ADD-R) property, it holds that
    \begin{equation}\label{igualdades}
\psi^1_i(N,c,\mathcal{B})=\underset{q : B_q\in \mathcal{B}^i}{\sum}\psi^1_{i(q)}(\bar{N},\bar{c},\bar{\mathcal{B}}) \mbox{ and }\psi^2_i(N,c,\mathcal{B})=\underset{q : B_q\in \mathcal{B}^i}{\sum}\psi^2_{i(q)}(\bar{N},\bar{c},\bar{\mathcal{B}}),
\end{equation}
where $(\bar{N},\bar{c},\bar{\mathcal{B}})$ is a game with a priori unions such that $(\bar{N},\bar{c})$ is the game specified in (\ref{cbar}) and $\bar{\mathcal{B}}$ is the partition in (\ref{Bbar}). 

Since  $\psi^1$ and $\psi^2$ are, in particular, coalitional Shapley values in the sense of \cite{vazquez1997owen} satisfying  (BC-CC) and (CQ) properties, we have ensured 
\begin{equation*}
\psi^1_{i(q)}(\bar{N},\bar{c},\bar{\mathcal{B}})=\psi^2_{i(q)}(\bar{N},\bar{c},\bar{\mathcal{B}})=Ow_{i(q)}(\bar{N},\bar{c},\bar{\mathcal{B}})
\end{equation*}
for all $i\in N$, where $i(q)$ is the representative of $i$ in $B_q\in \mathcal{B}$. The equalities in (\ref{igualdades}) and the assumption  $\psi^1_i(N,c,\mathcal{B})\neq \psi^2_i(N,c,\mathcal{B})$ lead directly to a contradiction. Then, the uniqueness is guaranteed.
    \item[(b)] \underline{Existence}. Now, we  prove the existence of a unique configurational Shapley value satisfying (ADD-R), (BC-CC) and (CQ) properties. 
    
    First, for the trivial configuration, the configuration value coincides with the Shapley value and is therefore a configurational Shapley value. \cite{albizuri2006configuration} proved that the configuration value satisfies (ADD-R), and Theorem \ref{propBC} and Theorem \ref{propCQ} ensure that the configuration value satisfies (BC-CC) and (CQ) properties.
    \end{itemize}
This concludes the proof.
\end{proof}

\section{The configuration value for airport games}\label{exact}
This section focuses on a new perspective of cooperation in airport settings based on the code-sharing policies. In particular, we propose a new allocation method for determining aircraft fees based on the configuration value for games with coalition configurations, which admits a polynomial expression. 

Let $(N,c)$ be an airport game and take $\mathcal{A}=\{1,\dots,A\}$ the set of airlines operating at the airport.  Take  $\mathcal{B}=\{B_1,\cdots,B_A\}$ the coalition configuration associated with the movements of planes in $N$, where $B_a$ formally corresponds to the subset of those movements operated by airline $a\in \mathcal{A}$. From now on, we call  $(N,c,\mathcal{B})$ the associated airport game with coalition configuration. \textcolor{black}{For each movement  $i\in N$, recall that $\mathcal{B}^i=\{B_a\in\mathcal{B}\ : \  i\in B_a\}$. In this context, $\mathcal{B}^i$ represents the subset  of airlines in $\mathcal{A}$ that operate movement $i$. In a code-sharing scenario, a flight may be operated by more than one airline. Thus, if movement $i$ is associated with two (or more) airlines, it directly holds that $|\mathcal{B}^i|> 1$.}

\textcolor{black}{The following result establishes a simple expression of the configuration value  for an airport game with coalition configuration $(N,c,\mathcal{B})$ which only depends on the parameters of the model. The exact expression of the configuration value} extends the ones provides for the Shapley value \citep{littlechild1973simple} and for  the Owen value \citep{vazquez1997owen} for airport games, respectively.

\begin{theorem}\label{theoremCV}
Let $(N,c,\mathcal{B})$ be an airport game with coalition configuration and $i\in N$ such that it corresponds to a movement of type $\tau(i)\in \{1,\dots,|{\cal T}|\}$ operated by a set of airlines specified by $\mathcal{B}^i$. Then, the configuration value of this movement is
	\begin{equation*}
	CV_{i}(N,c,\mathcal{B})=\underset{a : B_a\in \mathcal{B}^i}{\sum}\sum_{t=1}^{\tau(i)}\frac{c_{t}-c_{t-1}}{{|\mathcal{A}_{\geq t}|}{|N^{a}_{\geq t}|}}
	\end{equation*}
	being $c_0=0$, $N_{\geq t}^a=\cup_{\tau=t}^{|\cal T|}N_{\tau}\cap B_a$ the set of movements of type $t$ or larger of airline $a$ and $\mathcal{A}_{\geq t}=\{a\in \{1,\dots,A\}: N_{\geq t}^a\neq \emptyset\}$ the set of airlines with movements of
	type $t$ or larger.
\end{theorem}
\begin{proof}
  Take $(N, c, \mathcal{B})$  an airport game with a coalition configuration, and let $i \in N$ correspond to a movement of type $\tau(i) \in \{1, \dots, |{\cal T}|\}$ operated by the set of airlines indicated by $\mathcal{B}^i$. According to the equalities in (\ref{CV_OW}), the configuration value for any airport game with coalition configuration $(N,c,\mathcal{B})$ satisfies\vspace{0.2 cm}
  \begin{equation*}
CV_i(N,c,\mathcal{B})=\underset{a : B_a\in \mathcal{B}^i}{\sum}CV_{i(a)}(\bar{N},\bar{c},\bar{\mathcal{B}})=\underset{a : B_a\in \mathcal{B}^i}{\sum}Ow_{i(a)}(\bar{N},\bar{c},\bar{\mathcal{B}}),\vspace{0.2 cm}
\end{equation*}
where $(\bar{N},\bar{c},\bar{\mathcal{B}})$ represents the  game with a priori unions associated to $(N,c,\mathcal{B})$, specified by the TU game $(\bar{N},\bar{c})$, given in (\ref{cbar}), and $\bar{\mathcal{B}}$  the partition in (\ref{Bbar}).

In this case,  the Owen value for the airport game $(\bar{N},\bar{c},\bar{\mathcal{B}})$ associated to $(N,c,\mathcal{B})$, for a movement $i(a)$ in $\bar{N}$ (the representative of movement $i\in N$ in airline $a$), is given by\vspace{0.2 cm}
\begin{equation*}\label{owen2}
	Ow_{i(a)}(\bar{N},\bar{c},\bar{\mathcal{B}})=\sum_{t=1}^{\tau(i(a))}\frac{c_{t}-c_{t-1}}{{|\mathcal{A}_{\geq t}|}\ {|N^{a}_{\geq t}|}}=CV_{i(a)}(\bar{N},\bar{c},\bar{\mathcal{B}}).\vspace{0.2 cm}\end{equation*}
Therefore, as 
$Ow_{i}(\bar{N},\bar{c},\bar{\mathcal{B}})=CV_{i}(\bar{N},\bar{c},\bar{\mathcal{B}})$ and $\tau(i)=\tau(i(a))$ for any movement $i\in \bar{N}$ operated by an airline $a\in \{1,\cdots, A\}$, we have that
  \begin{equation*}
CV_i(N,c,\mathcal{B})=\underset{a : B_a\in \mathcal{B}^i}{\sum}CV_{i(a)}(\bar{N},\bar{c},\bar{\mathcal{B}})=\underset{a : B_a\in \mathcal{B}^i}{\sum}\sum_{t=1}^{\tau(i)}\frac{c_{t}-c_{t-1}}{{|\mathcal{A}_{\geq t}|}\ {|N^{a}_{\geq t}|}},
\end{equation*}
which concludes the proof.
\end{proof}

This distribution of costs has the following natural interpretation also in this new context, in line with ones justified for the Shapley value and the Owen value for airport games. The cost of the first portion of the runway, the cost $c_1$, which is the cost needed by all movements using the airport, is divided among all airlines involved, and within each airline, among all movements equally. The cost of the second part of the runway, $c_2-c_1$, is equally divided  among those airlines with movements of type greater than or equal to $2$ and, again, equally between movements that are of type $2,3,...,|\cal{T}|$ in each of these airlines. It continues progressively until the last tranche. 
For each movement of an aircraft operated by a set of airlines, the configuration value assigns to it the sum of the costs associated to such movement type across the different airlines operating it.

The following example illustrates the determination of the configuration value for an airport game with coalition configuration.
\begin{example}\label{ex0}Consider an airport situation involving four movements operated by three airlines. Movements 1, 3, and 4 are operated by airlines 1, 2, and 3, respectively. Movement 2 is jointly operated by airlines 2 and 3.  Movement 1 has an associated cost of 10 units (type 1), while movements 2 and 4 have a cost of 20 units each (type 2). Movement 3 incurs a cost of 22 units (type 3). 

We model this  situation as an airport problem with $N=\{1,2,3,4\}$, being the set of types of aircraft operating in the airport $\mathcal{T}=\{1,2,3\}$, with costs $c_1=10$, $c_2=20$ and $c_3=22$, respectively. Besides, the set of airlines that operate is specified by $\mathcal{A}=\{a_1,a_2,a_3\}$ and thus, the coalition configuration is $\mathcal{B}=\{B_{a_1},B_{a_2},B_{a_3}\}$, where $B_{a_1}=\{1\}$, $B_{a_2}=\{2,3\}$ and $B_{a_3}=\{2,4\}$. {The airlines that operate movements of type $1$ (or larger) are  $\mathcal{A}_{\geq 1}=\{a_1,a_2,a_3\}$; those that operate movements of type 2 (or larger) are given by $\mathcal{A}_{\geq 2}=\{a_2,a_3\}$ and those operating movement of type $3$ are $\mathcal{A}_{\geq 3}=\{a_2\}$. }Using these parameters, $(N,c,\mathcal{B})$ represents the associated airport game with coalition configuration, where $(N,c)$ is the airport game specified in Expression (\ref{airportgame}). 

\begin{figure}[h!]
    \centering
\begin{tikzpicture}[thick]
  \node at (-1.5,0.75) {$Type$};
  \node at (9.8,0.275) {$i$};
  \node at (10.5,0.25) {$\tau(i)$};
  \node at (11.4,0.275) {$\mathcal{B}^i$};
    \draw[-] (9.5,0.6) -- (11.8,0.6);
  \draw[-] (9.5,0) -- (11.8,0);
  \draw[-] (9.5,-2.6) -- (11.8,-2.6);

\node at (9.8,-0.4) {$1$};
\node at (9.8,-1) {$2$};
\node at (9.8,-1.6) {$3$};
\node at (9.8,-2.2) {$4$};

\node at (10.5,-0.4) {$1$};
\node at (10.5,-1) {$2$};
\node at (10.5,-1.6) {$3$};
\node at (10.5,-2.2) {$2$};

\node at (11.4,-0.4) {$a_1$};
\node at (11.45,-1) {$a_2, a_3$};
\node at (11.4,-1.6) {$a_2$};
\node at (11.4,-2.2) {$a_3$};

  \draw[-] (0,0) -- (4,0);
  \draw[-] (0,0.15) -- (0,-0.15);
  \draw[-] (4,0.15) -- (4,-0.15);
  \node at (-1.5,0) {$\tau=1$};
  \node at (2,0.3) {$c_1$};

   \draw[-] (0,-1) -- (8,-1);
   \draw[-] (0,-0.85) -- (0,-1.15);
   \draw[-] (8,-0.85) -- (8,-1.15);
   \node at (-1.5,-1) {$\tau=2$};
   \node at (4,-0.7) {$c_2$};

   \draw[-] (0,-2) -- (8.8,-2);
   \draw[-] (0,-1.85) -- (0,-2.15);
   \draw[-] (8.8,-1.85) -- (8.8,-2.15);
   \node at (-1.5,-2) {$\tau=3$};
   \node at (4.4,-1.7) {$c_3$};

   \draw[line width=6pt] (0,-3) -- (8.8,-3);
\draw[line width=2pt,color=white] (0.05,-3) -- (8.75,-3);
   \draw[-] (0,-1.85) -- (0,-2.15);
   \draw[-] (8.8,-1.85) -- (8.8,-2.15);
     \draw[line width=6pt] (-0.05,-3) -- (0.05,-3);
    \draw[line width=6pt] (3.95,-3) -- (4.05,-3);
     \draw[line width=6pt] (7.95,-3) -- (8.05,-3);
    \draw[line width=6pt] (8.75,-3) -- (8.85,-3);
   \node at (-1.5,-3) {Runway};
      \node at (0,-3.46) { $0$};
      \node at (4,-3.46) { $10$};
 \node at (8,-3.46) { $20$};
    \node at (8.8,-3.46) { $22$};

\end{tikzpicture}\caption{Scheme of the runway by the three types of planes.}
    \label{fig:runway}
\end{figure}
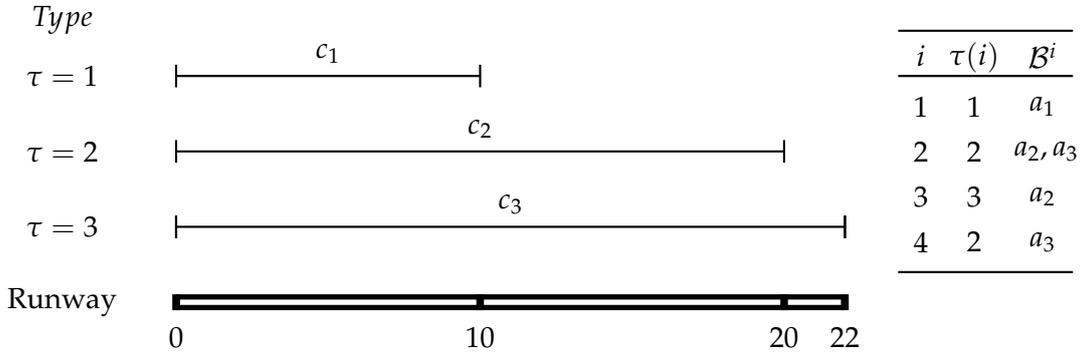

Below, we compute the configuration value.  First, take movement $i=1$. We have that $\mathcal{B}^1=\{B_{a_1}\}$. Hence, since $\tau(1)=1$,    $|N^{a_1}_{\geq 1}|=1$, we have that\vspace{0.1 cm}
\begin{equation*}
    CV_{1}(N,c,\mathcal{B})=\frac{10-0}{3\cdot 1}=\frac{10}{3}.\vspace{0.1 cm}
\end{equation*}
For movement $i=2$, we have $\mathcal{B}^2=\{B_{a_2},B_{a_3}\}$. Since $\tau(2)=2$, we have that  $|N^{a_2}_{\geq 1}|=|N^{a_2}_{\geq 2}|=|N^{a_3}_{\geq 1}|=|N^{a_3}_{\geq 2}|=2$, and thus,
\begin{equation*}
 CV_{2}(N,c,\mathcal{B})=\left(\frac{10-0}{3\cdot 2}+\frac{20-10}{2\cdot 2}\right)+\left(\frac{10-0}{3\cdot 2}+\frac{20-10}{2\cdot 2}\right)=\frac{50}{6}.
\end{equation*}
For movement $i=3$, $\mathcal{B}^3=\{B_{a_2}\}$. Since $\tau(3)=3$, we have that  $|N^{a_2}_{\geq 1}|=1$, $|N^{a_2}_{\geq 2}|=2$ and $|N^{a_2}_{\geq 3}|=1$, and thus,
\begin{equation*}
    CV_{3}(N,c,\mathcal{B})=\left(\frac{10-0}{3\cdot 2}+\frac{20-10}{2\cdot 2}\right)+\left(\frac{22-20}{1\cdot 1}\right)=\frac{37}{6}.
\end{equation*}
Finally, for movement $i=4$, we have $\mathcal{B}^4=\{B_{a_3}\}$. Since $\tau(4)=2$, we  have that $|N^{a_3}_{\geq 1}|=2$ and  $|N^{a_3}_{\geq 2}|=2$, and thus,
\begin{equation*}
CV_{4}(N,c,\mathcal{B})=\frac{10-0}{3\cdot 2}+\frac{20-10}{2\cdot 2}=\frac{25}{6}.
\end{equation*}
Then, the configuration value for $(N,c,\mathcal{B})$ is $CV(N,c,\mathcal{B})=(10/3,50/6,37/6,25/6)$. In view of the resulting allocation fee, we check that, according to the configuration value, movement 3 pays less than movement 2, even though it corresponds to a movement of a larger type.
\end{example}

{\color{black}Similar to the reasoning considered for the Shapley and the Owen value for airport games, when using the configuration value to distribute operating costs in airport situations, the total fee paid by each airline depends only on the types of aircraft of that airline making movements at the airport and not on the number of aircraft of the airline. For airlines that use the airport frequently (with a high number of movements), the total fees can be spread over more movements, resulting in a lower fee per movement. Conversely, airlines that use the airport less frequently will incur higher fees per movement. Furthermore, the total fee of an airline remains unchanged if it increases its movements at the airport using aircraft that are smaller or equivalent in size to those it already operates there.}

\textcolor{black}{The axioms originally considered in Theorem \ref{thalb} now serve as fairness and consistency principles in the context of airport cost allocation, particularly under the presence of a coalition configuration. Specifically, efficiency (EFF) ensures that the total cost is fully allocated among all aircraft movements;  null player property (NPP) requires that any aircraft movement which does not increase the cost of any coalition receives an allocation fee equal to zero;  linearity (L) guarantees that, when an airport game can be expressed as a linear combination of simpler airport games, the allocation fee is computed as the corresponding linear combination of the fees from such problems;  anonymity (A) implies that the roles or positions in the aircraft configuration has no influence in the final allocation fee;  coalitional symmetry (CS) guarantees that, if there exist a pair of symmetric airline alliances (i.e. with non-cost impact when removing each one on the remaining), both receive the same total allocation fee; and merger (M) implies that when two "double" movements (essentially identical in behavior and coalition roles) exist, removing one does not affect others' allocation fees.}

\textcolor{black}{However, the fact that the configuration value is a configurational Shapley value allows new interpretations of the resulting allocations. The first is that if each airline operates a single movement, the configuration value of the corresponding airport game is the Shapley value  given in Equation (\ref{shapley}).  Furthermore, the properties that characterize the configuration value for  games with coalition configuration in Theorem \ref{ThCV} allow for a reinterpretation of the aircraft fees obtained by applying it to airport games.} First, the additivity in representatives property (ADD-R) ensures that the fee assigned to a movement by the configuration value equals the sum of the fees corresponding to the different airlines involved in operating it. {The property of balanced contributions for coalition configurations (BC-CC) ensures that if two movements, $i$ and $j$, are operated by the same subset of airlines, then the difference of the configuration value of $i$ and $j$ is equal to the sum of the differences of the allocations for each movement when the other leaves each airline that operates it.} Finally, the coalitional-quotient game (CQ) property implies that, in airport games, it makes no difference if the airport authorities calculate fees based on movements or by airline. This property can be viewed as a generalization of the quotient game property associated with the Owen value, adapted to scenarios where a coalition configuration exists.

\textcolor{black}{In the following, we briefly introduce a final comment on the allocation fee of each airline. Let $(N,c,\mathcal{B})$ be an airport game with coalition configuration such that each $i\in N$ identifies a movement of type $\tau(i)\in \{1,\dots,|{\cal T}|\}$ operated by the set of airlines in $\mathcal{B}^i$. Then, if $(\bar{N},\bar{c},\bar{\mathcal{B}})$ denotes the game with a priori unions given by  the TU game $(\bar{N},\bar{c})$ in (\ref{cbar}) and the partition $\bar{\mathcal{B}}$ in (\ref{Bbar}), the configuration value assigns a total fee for each airline $a$ equal to\vspace{0.2 cm}
	\begin{equation}\label{CVa}
	 TF^{CV}_{a}(N,c,\mathcal{B})=\sum_{t\in \mathcal{T}: \ {N}_t\cap {B}_a\neq \emptyset}|{N}_t\cap {B}_a|\cdot CV_{a,t}(\bar{N},\bar{c},\bar{\mathcal{B}}),\vspace{0.2 cm}
	 \end{equation}
where $CV_{a,t}(\bar{N},\bar{c},\bar{\mathcal{B}})=\sum_{k=1}^{t}\frac{c_{k}-c_{k-1}}{|\mathcal{A}_{\geq k}||N^{a}_{\geq k}|}$ for any $t\in \mathcal{T}$. Such amount represents the portion of the allocation fee specified by the configuration value for the representative of a movement in $N$ of type $t\in \{1,\dots,|{\cal T}|\}$ operated by an airline $a\in \{1,\dots,A\}$.}

Below, we illustrate the obtaining of the allocation fee per airline based on the configuration value.

\begin{example}
We revisit the airport game with coalition configuration $(N,c,\mathcal{B})$ in Example \ref{ex0}. For each airline $a\in \mathcal{A}$, the total fee based on the configuration value can be determined using (\ref{CVa}). For airline $a_1$, we have
\begin{equation*}
    TF^{CV}_{a_1}(N,c,\mathcal{B})=\frac{10-0}{3\cdot 1}=\frac{10}{3}.
\end{equation*}
For the case of airline $a_2$,
\begin{equation*}
    TF^{CV}_{a_2}(N,c,\mathcal{B})=\left(\frac{10-0}{3\cdot 2}+\frac{20-10}{2\cdot 2}\right)+\left(\frac{10-0}{3\cdot 2}+\frac{20-10}{2\cdot 2}\right)+\left(\frac{22-20}{1\cdot 1}\right)=\frac{31}{3}.
\end{equation*}
Finally, when considering airline $a_3$, we have
\begin{equation*}
    TF^{CV}_{a_3}(N,c,\mathcal{B})=\left(\frac{10-0}{3\cdot 2}+\frac{20-10}{2\cdot 2}\right)+\left(\frac{10-0}{3\cdot 2}+\frac{20-10}{2\cdot 2}\right)=\frac{25}{3}.
\end{equation*}
Then, the fee for airlines specified by the configuration value for $(N,c,\mathcal{B})$ is $TF^{CV}(N,c,\mathcal{B})=(10/3,31/3,25/3)$.
\end{example}

Take a subset of airlines  $\mathcal{M}\subseteq \mathcal{A}$ that forms a new airline  $a^{\star}$. Thus, it results $\mathcal{A}^{\ast}=\{a^{\ast},\mathcal{A}\setminus \mathcal{M}\}$ as a new set of airlines and $(N,c,\mathcal{B}^{\ast})$  as the airport game with coalition configuration that corresponds with the same set of movements and the original airport game. However, the coalition configuration $\mathcal{B}^{\ast}$ is now given by $\mathcal{B}^{\ast}=\{B_{a^{\ast}},\{B_{a^{\prime}}: a^{\prime}\in \mathcal{A}\setminus\mathcal{M}\}\}$, with $B_{a^{\ast}}=\underset{\alpha\in \mathcal{M}}{\cup}B_{\alpha}$.

Below, we formalize a result on the profit of airlines in  our context, aligned with the result proved for the case of the Owen value for airport games in \cite{vazquez1997owen}.

\begin{proposition}Let $(N, c, \mathcal{B})$ be an airport game with coalition configuration and consider a subset $\mathcal{M}$ of airlines in $\mathcal{A}$ that decides to form a new airline $a^{\ast}$. If $(N, c, \mathcal{B}^{\ast})$ is the  airport game with coalition configuration resulting from the merging, it holds that $TF^{CV}_{a^{\ast}}(N,c,\mathcal{B}^{\ast})\leq\sum_{\alpha\in \mathcal{M}}TF^{CV}_{\alpha}(N,c,\mathcal{B})$.  
\end{proposition}
\begin{proof}Take $(N, c, \mathcal{B})$ an airport game with coalition configuration and denote by  $\mathcal{M}$ the  subset of airlines in $\mathcal{A}$ merge into an airline $a^{\ast}$. Consider $(N,c,\mathcal{B}^{\ast})$  the airport game with coalition configuration resulting from such airline. Thus, by {Expression (\ref{CVa}),} by the relation between the configuration value and the Owen value for games with a priori unions and by using Theorem 1 in \cite{vazquez1997owen}, it holds that
\begin{equation*}
\begin{split}
TF^{CV}_{a^\ast}(N,c,\mathcal{B}^{\ast})&=\sum_{t\in \mathcal{T}: \ {N}_t\cap {B}_{a^\ast}\neq \emptyset}|{N}_t\cap {B}_{a^\ast}|\cdot CV_{a^\ast,t}(\bar{N},\bar{c},\bar{\mathcal{B}}^\ast)\\
&=\sum_{t\in \mathcal{T}: \ {N}_t\cap {B}_{a^\ast}\neq \emptyset}|{N}_t\cap {B}_{a^\ast}|\cdot Ow_{a^\ast,t}(\bar{N},\bar{c},\bar{\mathcal{B}}^\ast)\\
&=TF^{Ow}_{a^\ast}(\bar{N},\bar{c},\bar{\mathcal{B}}^{\ast})\\
& \leq\sum_{\alpha\in \mathcal{M}}TF^{Ow}_{\alpha}(\bar{N},\bar{c},\bar{\mathcal{B}})\\
&=\sum_{\alpha\in \mathcal{M}}TF^{CV}_{\alpha}(N,c,\mathcal{B}),
\end{split}
\end{equation*}
where $\bar{\mathcal{B}}^\ast$ is the partition of the representatives of $\bar{N}$ associated with ${\mathcal{B}}^\ast$, and $TF^{Ow}_{a}(N,c,P)$ denotes the allocation fee specified by the Owen value for a game with a priori unions $(N,c,P)$ for  an airline $a\in\mathcal{A}$.\end{proof}

\section{An application: the case of a Spanish airport}\label{sec:results}

{In this section, we present a comparative analysis of fees for the aircraft operations at the airport of  Santiago de Compostela - Rosal\'ia de Castro (IATA code: SCQ) in Galicia (Spain) on two non-working days using game-theoretic solutions as allocation mechanisms. From now on, when we refer to an operation at an airport, we mean only the take-off or landing of an aircraft.}

{The following subsections present the numerical results. In Subsection~\ref{sec:51}, we first describe the elements that define the associated airport problem with coalition configuration and we determine the configuration value. Subsection~\ref{sec:comp} compares these results with those obtained with the Shapley value and the Owen value.}

\subsection{A  new system of aircraft fees  based on the configuration value}\label{sec:51}

We consider the real dataset obtained from 131 movements of aircraft operating at the airport on the 23rd and 24th October 2024.  Table \ref{airplanes} lists the nine types of aircraft movements and the take-off weight of each type of plane (in tons) which undoubtedly influence its operating costs.

\begin{table}[!h]
	\begin{center}
		\resizebox{0.9\textwidth}{!}{	
		\begin{tabular}{|p{0.9cm}|p{1cm}|p{4.8cm}|p{2cm}||p{2.5cm}|p{2cm}|}	\hline
Type &  \multicolumn{2}{l|}{Aircraft Type}  &
Take-off Weight (t) &
AENA's fees (euros p.m.) & Number of movements\\\hline\hline
1 &LJ45 & Learjet 45 &9.752&43.3412 & 2\\
2 & C650 & Cessna Citation VI &10.660&47.3766& 2\\
3 & C68A & Cessna Citation Sovereign &13.744&61.0830 & 1\\
4 & CL35 & Bombardier Challenger 350 &18.430&81.9091 & 2\\
5 & CRJ2 & Bombardier CRJ200 &24.041&106.8463 & 7\\
6 &  A320 & Airbus A320       & 78.000&346.6584 & 50\\
7 & A20N & Airbus A320neo&79.000&351.1027 & 16\\
8 & B738 & Boeing 737-800&79.015&351.1694 & 30\\
9 & B38M & Boeing 737 MAX 8     &	82.190	&365.2801
 & 21\\\hline
\end{tabular}}\caption{Types of airplanes operating at the airport  Santiago de Compostela - Rosal\'ia de Castro and AENA's fees per movement.} \label{airplanes}
	\end{center}
\end{table}

  Table \ref{airplanes} also shows the cost for each type of movement   according to the fee specifically designed  for operations at the airport  Santiago de Compostela - Rosal\'ia de Castro by the Spanish regulatory agency \emph{Aeropuertos Espa\~noles y Navegaci\'on A\'erea (AENA)}.\footnote{More information on the operation and management of Spanish airports can be found at \url{https://www.aena.es/}. In particular, the monthly fees applicable to aircraft operations in Spain are detailed at \url{https://www.aena.es/es/aerolineas/tarifas.html} Data for October 2024 are available on request.} It should be noted that these fees depend on the take-off weight associated with each type of plane. The table also displays the number of movements  present at the airport for each type of aircraft. Taking into account both amounts, the total cost fee to be collected by the authorities managing all these aircraft operations on the runway at Santiago de Compostela airport during the  period considered is 42310.79 euros. \textcolor{black}{We can naturally assume that airport managers have to collect such amount for operations in the considered period, regardless of the number or type of aircraft they operate.} Notice that this is the amount to be paid if there only operates  one movement and it corresponds to  the largest type. As AENA details in its official documents, AENA's fees for each type of aircraft are proportional to its take-off weight as last column of Table \ref{airplanes} shows.
  
  We model the problem of distributing this overall cost as an airport problem with coalition configuration. Firstly, the cost associated with each aircraft type is the proportion of the total cost corresponding to its take-off weight.  They are  displayed in Table 3. 
  
  
  

\begin{table}[h!]
	\begin{center}\resizebox{0.675\textwidth}{!}{	
		\begin{tabular}{|l|r|r|r|r|r|r|r|r|r|}\hline
   Type  $\tau$ & 1 & 2 & 3 & 4 & 5 \\\hline
$c_{\tau}$ & 5020.255 & 5487.687 & 7075.307 & 9487.624 & 12376.124 \\\hline\hline
 Type $\tau$ &  6 & 7 & 8 & 9 &\\\hline
 $c_{\tau}$ &40153.807 & 40668.599 & 40676.321 & 42310.790 & \\\hline
		\end{tabular}}\caption{Distribution of operating costs per type of airplane.}\label{coststype}\end{center}
\end{table}

However, the code-sharing information of each movement now needs to be considered. Table \ref{airlines} enumerates the list of the airlines that operate under code-sharing agreements at the airport  Santiago de Compostela - Rosal\'ia de Castro during the period considered. 
\begin{table}[!ht]
	\begin{center}
		\resizebox{0.55\textwidth}{!}{	
		\begin{tabular}{|p{0.6cm}p{3.25cm}|p{0.6cm}p{3.25cm}|}	\hline
  $A_i$ & Airline & $A_i$ & Airline \\\hline\hline
$A_1$ & Ryanair  & $A_{11}$ & British Airways\\          
$A_2$ & Vueling & $A_{12}$ & Lauda Europe \\            
$A_3$ & Iberia & $A_{13}$ & World2Fly\\      
$A_4$ & easyJet & $A_{14}$ & Royal Air Maroc\\        
$A_5$ & Qatar Airways & $A_{15}$ & Japan Airlines \\ 
 $A_6$ & American Airlines  & $A_{16}$ & Iberia Regional\\ 
 $A_7$ & Avianca& $A_{17}$ & Unicair\\ 
 $A_8$ & JetStream  & $A_{18}$ & Jet Linx Aviation\\ 
 $A_9$ & VistaJet  & $A_{19}$ & NetJets Europe\\ 
 $A_{10}$ & Lufthansa & &\\\hline		
 \end{tabular}}\caption{List of airlines that operate at the airport.} \label{airlines}
	\end{center}
\end{table}
\newpage Thus, the organization of the 131 flights operated at the airport during the period considered is presented in Table \ref{dataSCQ}. Each column represents an airline (an element of the coalition configuration), while each row indicates, with a tick, the airlines that operate each subset of movements. For example, from the second row of the table, we check that Vueling and Iberia ($A_2$ and $A_3$) jointly operate 6 movements of planes A320 and 7 of planes A20N. Thus, for each possible group of airlines (each row of the table), we list the flights per type of aircraft operated for them.

All these elements allow to design the corresponding airport game with a coalition configuration and, thus, to determine the associated aircraft fee based on the configuration value. Last column of  Table \ref{dataSCQ} gives the distribution of the operating costs per movement specified by it, identifying the type of airplane and the airlines that operate it. The fees are given in euros.

\begin{landscape}
\begin{table}[!h]
	\begin{center}
		\resizebox{0.9\textwidth}{!}{	
		\begin{tabular}{|p{0.42cm}|p{0.42cm}|p{0.42cm}|p{0.42cm}|p{0.42cm}|p{0.42cm}|p{0.42cm}|p{0.42cm}|p{0.42cm}|p{0.5cm}|p{0.5cm}|p{0.5cm}|p{0.5cm}|p{0.5cm}|p{0.5cm}|p{0.5cm}|p{0.5cm}|p{0.5cm}|p{0.5cm}||p{1.65cm}|r|r|}	\hline
  \multirow{2}{*}{$A_1$} & \multirow{2}{*}{$A_2$} & \multirow{2}{*}{$A_3$} & \multirow{2}{*}{$A_4$} & \multirow{2}{*}{$A_5$}  & \multirow{2}{*}{$A_6$} & \multirow{2}{*}{$A_7$} & \multirow{2}{*}{$A_8$} & \multirow{2}{*}{$A_9$} & \multirow{2}{*}{$A_{10}$} & \multirow{2}{*}{$A_{11}$} & \multirow{2}{*}{$A_{12}$} & \multirow{2}{*}{$A_{13}$} & \multirow{2}{*}{$A_{14}$} & \multirow{2}{*}{$A_{15}$} & \multirow{2}{*}{$A_{16}$} & \multirow{2}{*}{$A_{17}$} & \multirow{2}{*}{$A_{18}$} & \multirow{2}{*}{$A_{19}$} & \multirow{2}{*}{Type} &  Number of & Configuration \\
    &  &  & & &  &  &  &  & & & & & & &  &  &  & &  & movements & value (p.m.)\\\hline\hline
\multirow{2}{*}{\checkmark} & & & & & & & & & & & & & & & & & & &B738 (8) & 30 & 54.318\\
 & & & & & & & & & & & & & & & & & & &  B38M (9) & 21 & 132.150\\\hdashline
  & \multirow{2}{*}{\checkmark}& \multirow{2}{*}{\checkmark}& & & & & & & & & & & & & & & & & A320 (6) & 21 & 106.540\\
  & & & & & & & & & & & & & & & & & & & A20N (7) & 1 &  116.836\\\hdashline
  & & & \multirow{2}{*}{\checkmark}& & & & & & & & & & & & & & & & A320 (6) & 3 & 542.208\\
  & & & & & & & & & & & & & & & & & & & A20N (7)& 2 & 567.947\\\hdashline
  & \multirow{2}{*}{\checkmark}& \multirow{2}{*}{\checkmark}& & \multirow{2}{*}{\checkmark}& & & & & & & & & & & & & & & A320 (6) & 9 & 235.637\\
  & & & & & & & & & & & & & & & & & & & A20N (7) & 2 & 254.513\\\hdashline
  & \multirow{2}{*}{\checkmark}& \multirow{2}{*}{\checkmark}& & \multirow{2}{*}{\checkmark}& \multirow{2}{*}{\checkmark}& \multirow{2}{*}{\checkmark}& & & & & & & & & & & & & A20N (7)& 2 &1218.331\\
  & & & & & & & & & & & & & & & & & & & A320 (6) & 1 &1165.136\\\hdashline
  & \multirow{2}{*}{\checkmark}& & & & & & & & & & & & & & & & & & A320 (6)& 5 & 48.858\\
  & & & & & & & & & & & & & & & & & & & CRJ2 (5)& 1 & 12.115\\\hdashline
  & & & & & & & \checkmark& & & & & & & & & & & & C650 (2) & 2 & 145.096\\\hdashline
  & & & & & & & & \checkmark& & & & & & & & & & & CL35 (4)& 2 &267.176\\\hdashline
  & & & & & & & & & \checkmark& & & & & & & & & & A20N (7) & 2 &1381.259\\\hdashline
  & \multirow{2}{*}{\checkmark}& \multirow{2}{*}{\checkmark}& & & & & & & & \multirow{2}{*}{\checkmark}& & & & & & & & & A320 (6)& 2 & 784.300\\
  & & & & & & & & & & & & & & & & & & & A20N (7)& 2 & 820.335\\\hdashline
  & & & & & & & & & & & \checkmark& & & & & & & & A320 (6) & 2 &1355.519\\\hdashline
  & \checkmark& \checkmark& & \checkmark& & \checkmark& & & & & & \checkmark& \checkmark& \checkmark& & & & & A320 (6) & 1 & 5593.165\\\hdashline
  & \multirow{2}{*}{\checkmark}& & & \multirow{2}{*}{\checkmark}& & & & & & & & & & & & & & & A320 (6)& 3 & 177.955\\
  & & & & & & & & & & & & & & & & & & & A20N (7) & 2 & 191.683\\\hdashline
  & \checkmark& & & & & & & & & & & & & & \checkmark& & & & CRJ2 (5) & 5 & 133.268\\\hdashline
  & & & & & & & & & & & & & & & & \checkmark& & & LJ45 (1) & 2 & 132.112\\\hdashline
  & & \multirow{2}{*}{\checkmark}& & & & & & & & & & & & & & & & & A320 (6) & 1 & 57.682\\
  & & & & & & & & & & & & & & & & & & & A20N (7) & 2 & 62.830\\\hdashline
  & \checkmark& \checkmark& & & \checkmark& \checkmark& & & & & & & \checkmark& & & & & & A20N (7) & 1 & 2035.813\\\hdashline
  & \checkmark& \checkmark& & & & \checkmark& & & & & & & & & & & & & A320 (6) & 1 & 493.831\\\hdashline
  & \checkmark& \checkmark& & \checkmark& \checkmark& \checkmark& & & & & & \checkmark& \checkmark& & & & \checkmark& & A320 (6) & 1 & 6135.373\\\hdashline
  & & & & & & & & & & & & & & & & & & \checkmark& C68A (3)& 1 & 383.582\\\hdashline
& & & & & & & & & & & & & & & \checkmark& & & & CRJ2 (5) & 1 & 121.153\\\hline
\end{tabular}}
		\caption{Configuration value per movement (p.m.).} \label{dataSCQ}
	\end{center}
\end{table}
\end{landscape}


We can draw some conclusions from the last column of Table \ref{dataSCQ}. Aircraft movements operated by airlines that use an airport occasionally face higher movement fees, while those operating frequently benefit from lower fees. For example, A320 aircraft operated solely by $A_{15}$ and $A_{18}$ incur significantly higher fees compared to those operated by $A_2$ and $A_3$, which have many movements. Smaller aircraft like the LJ45 also exhibit a fee per operation similar to that for B38M and higher than that for B738, both operated by  $A_1$ and corresponding to the two largest aircraft types. Since fees are charged per movement, airlines with frequent operations can distribute costs more effectively, leading to lower per-movement fees. However, cooperation among airlines affects cost allocation. For instance, $A_3$ (Iberia) collaborates widely, resulting in lower fees for A320 and A20N movements, whereas the same aircraft types operated by $A_4$, which does not operate under code-sharing agreements, face much higher fees.

Then, once the configuration value has been determined, we  calculate the fees associated with each of the airlines using Expression (\ref{CVa}), that are collected in Table \ref{airlinesfees}. Table \ref{dataSCQ_costs} in Appendix \ref{appA} of the Online Resource Section (ORS) shows the associated fees per airline and per movement whose sum is the configuration value.

\begin{table}[!ht]
	\begin{center}
		\resizebox{0.8\textwidth}{!}{	
		\begin{tabular}{|p{0.6cm}p{3.25cm}p{1.7cm}|p{0.6cm}p{3.25cm}p{1.7cm}|}	\hline
  $A_i$ & Airline & Total Fee & $A_i$ & Airline &Total Fee  \\\hline\hline
$A_1$ & Ryanair &4404.690 & $A_{11}$ & British Airways&	2762.518\\          
$A_2$ & Vueling&2762.502 & $A_{12}$ & Lauda Europe &2711.038\\            
$A_3$ & Iberia &2762.534& $A_{13}$ & World2Fly &2711.038\\      
$A_4$ & easyJet&	2762.518 & $A_{14}$ & Royal Air Maroc&2762.519\\        
$A_5$ & Qatar Airways&	2762.517& $A_{15}$ & Japan Airlines &2711.038\\ 
 $A_6$ & American Airlines &	2762.517  & $A_{16}$ & Iberia Regional & 726.918\\ 
 $A_7$ & Avianca&	2762.517& $A_{17}$ & Unicair & 264.224\\ 
 $A_8$ & JetStream &290.192& $A_{18}$ & Jet Linx Aviation& 2711.038\\ 
 $A_9$ & VistaJet &534.352& $A_{19}$ & NetJets Europe &383.582\\ 
 $A_{10}$ & Lufthansa &2762.518& & &\\\hline		
 \end{tabular}}\caption{Total fees for airlines based on the configuration value.} \label{airlinesfees}
	\end{center}
\end{table}

From these results, we make a few brief comments. Ryanair is the highest paying airline with 51 operations at the airport. On the other hand, Iberia Regional, VistaJet, NetJets Europe, JetStream and Unicair are the lowest paying airlines. The remaining airlines pay just over 2700 euros.

\subsection{A comparative analysis: the Owen value and the Shapley value}\label{sec:comp}

\cite{vazquez1997owen} also studied the effects of taking into account airline alliances in determining airport fees by considering  the Owen value of an airport game. To do this, we now look at the distribution of aircraft movements at Santiago de Compostela - Rosal\'ia de Castro airport according to the airline alliance that owns the aircraft operating the flight. In this way, we highlight the cases of  the airline alliances One World and Star Alliance. Table \ref{airlinespartition} indicates whether each airline belongs to one of the two alliances mentioned above or whether it operates independently. This results into the partition to be taken into account in determining the new system of aircraft fees.

\begin{table}[!h]
	\begin{center}
		\resizebox{0.88\textwidth}{!}{	
		\begin{tabular}{|p{0.6cm}|p{2.8cm}|p{10cm}|}	\hline
  $P_i$ & Airline alliance & Airlines\\\hline\hline
$P_1$ & One World & Vueling, Iberia, Iberia Regional, Qatar Airways, American Airlines, Avianca, British Airways, World2Fly, Royal Air Maroc, Japan Airlines, Jet Linx Aviation\\\hdashline[0.5pt/5pt]
$P_2$ & Star Alliance &  Lufthansa\\\hdashline[0.5pt/5pt]
$P_3$ &- & Ryanair\\\hdashline[0.5pt/5pt]
$P_4$ &- & easyJet\\\hdashline[0.5pt/5pt]
$P_5$ &- & JetStream\\\hdashline[0.5pt/5pt]
$P_6$ &- & VistaJet\\\hdashline[0.5pt/5pt]
$P_7$ &- & Lauda Europe\\\hdashline[0.5pt/5pt]
$P_{8}$ &- & Unicair\\\hdashline[0.5pt/5pt]
$P_{9}$ & -& NetJets Europe\\\hline		\end{tabular}
		}
		\caption{List of airline alliances that operate at the airport of Santiago de Compostela.} \label{airlinespartition}
	\end{center}
\end{table}

Table \ref{alliances} provides a description of the airline alliances at the airport, including the grouping of movements according to these alliances, as well as the Owen value for aircraft movements (calculated using Expression (\ref{owen})), which is distinguished by type and airline. The resulting fees per movement (p.m) are given in euros.\vspace{0.25 cm}
\begin{table}[h!]
	\begin{center}\small\resizebox{0.7\textwidth}{!}{	
		\begin{tabular}{|p{3cm}||p{2.25cm}p{2cm}|p{2cm}|}\hline
  Airline alliance & Aircraft Type & Number of movements & Owen value (p.m.)\\\hline\hline
OneWorld	&	A320 (6)	&	45 & 125.947\\
	&	A20N (7)	&	12 &   136.672\\
	&	CRJ2 (5)	&	7  & 28.481\\\hdashline[0.5pt/5pt]
Ryanair	&	B738 (8)	&	30 &147.348\\
	&	B38M (9)	&	21 & 225.180\\\hdashline[0.5pt/5pt]
easyJet	&	A320 (6)	&	3 & 1475.666\\
	&	A20N (7)	&	2 & 1540.015\\\hdashline[0.5pt/5pt]
JetStream	&	C650 (2)	&	2 &  308.118\\\hdashline[0.5pt/5pt]
VistaJet	&	CL35 (4)	&	2 &  622.545\\\hdashline[0.5pt/5pt]
Lufthansa	&	A20N (7)	&	2 & 3753.513\\\hdashline[0.5pt/5pt]
Lauda Europe	&	A320 (6)	&	2 & 3689.164\\\hdashline[0.5pt/5pt]
Unicair	&	LJ45 (1) &	2 &   278.903\\\hdashline[0.5pt/5pt]
NetJets Europe	&	C68A (3)	&	1 & 843.038\\
\hline
		\end{tabular}}\caption{Owen value per movement (p.m.).}\label{alliances}\end{center}
\end{table}

In view of the results, we make a few brief comments.  The movements of A320 aircraft operated by OneWorld would generally incur lower charges under the Owen value than under the configuration value. The exceptions are those movements operated jointly by $A_2$ and $A_3$ and those operated exclusively by $A_2$ and $A_3$. Particularly noteworthy are the movements of Ryanair, for which the Owen value assigns significantly higher costs compared to the configuration value. Regarding other airlines, it is worth noting that major companies such as EasyJet, Lauda Europe, or Lufthansa, which have fewer movements at the airport, would also prefer the configuration value as a cost allocation method. The same applies to companies such as JetStream (C650), VistaJet (CL35), Unicair (LJ45), and NetJets Europe (C68A), which operate  the smallest aircraft types. Broadly speaking, we can conclude that airline alliances would prefer using the configuration value to determine their fees. Meanwhile, those airline alliances that are more active, have a greater presence and operate larger aircraft tend to benefit more from the Owen value.

Finally, we  obtain  the cost distribution using the Shapley value for airport games \citep{littlechild1973simple}. Such approach overlooks the reality of alliances and code-sharing agreements among airlines for airport operations. Last column of Table \ref{distributionmovements} specifies the corresponding Shapley value per movement at the airport.

\begin{table}[h!]
	\begin{center}\small\resizebox{0.425\textwidth}{!}{	
		\begin{tabular}{|p{1.5cm}|p{2cm}|p{2cm}|}\hline
   Aircraft Type & Number of movements & Shapley value (p.m.)\\\hline\hline
LJ45	&2 &  38.323\\
C650	&2 & 41.946\\
C68A	&1 & 54.447\\
CL35	&2 &73.592 \\
CRJ2	&7 & 96.887\\
A320	&50 &334.303\\
A20N	&16 & 341.986\\
B738	&30 & 342.138\\
B38M &	21 & 419.969\\\hline
		\end{tabular}}\caption{Shapley value per  movement (p.m.).}\label{distributionmovements}\end{center}
\end{table}

Finally, we globally compare the allocations obtained under the cooperative game approach, using the Shapley value, the Owen value, and the configuration value, with the allocation fees that AENA collects for each type of movement. Table \ref{aggregatefees} presents the average fee per type of movement across the four scenarios considered. Note that under the Shapley value, aircraft of the same type are assigned the same fee. This is not the case for the allocations derived using the Owen value and the configuration value, as these depend on the number of airlines operating each movement, even if the aircraft are of the same type.

\begin{table}[h!]
	\begin{center}\small\resizebox{0.8\textwidth}{!}{	
		\begin{tabular}{|p{1.5cm}|p{2cm}||p{2.25cm}|p{1.5cm}|p{1.45cm}|p{1.45cm}|}\hline
   Aircraft Type & Number of movements & Configuration value & Owen value & Shapley value & AENA\\\hline\hline
LJ45 (1)	&2 & 132.112&278.903&38.323&43.341\\
C650 (2)	&2 &145.096&308.118&41.946&47.377\\
C68A (3)	&1 & 383.582&843.038&54.447&61.083\\
CL35 (4)	&2 & 267.176&622.545&73.592&81.909\\
CRJ2 (5)	&7 & 114.230&28.481&96.887&106.846\\
A320 (6)	&50 & 489.754&349.459&334.303&346.658\\
A20N (7)	&16 &696.653&764.195&341.986&351.103\\
B738 (8)	&30 &54.318&147.348&342.138&351.169\\
B38M (9)    &21 & 132.150&225.180&419.969&365.280\\\hline
		\end{tabular}}\caption{Average fees per movement (p.m.) for each  type of airplane.}\label{aggregatefees}\end{center}
\end{table}

Based on the numerical results, it appears that the fees based on the Shapley value are usually, on average and for each type of aircraft, slightly lower than those collected by AENA. The only exception is for type 9 aircraft movements, where the Shapley value assigns a larger amount. When considering the Owen value, the resulting fees generally increase for airlines with fewer movements, while favoring airlines with higher levels of activity. Even so, we observe that the A320 and A20N aircraft movements of EasyJet and Lufthansa (which have limited activity at the airport) raise the average cost of these aircraft types. In a sense, the configuration value minimizes this effect by reducing the fees associated with small aircraft movements and penalizes   the movements operated under code-sharing  by any airline with limited presence at the airport. Analogous comments can be inferred from the analysis of the total fees per airline alliance, which are shown in Table \ref{aggregateairlines}. We can see that with respect to the Shapley value, the Owen value favors large alliances or airlines with a large number of movements by assigning them lower fees. This effect is offset by the configuration value.

\begin{table}[h!]
	\begin{center}\small\resizebox{0.8\textwidth}{!}{	
		\begin{tabular}{|p{2.5cm}||p{2.25cm}|p{2cm}|p{2cm}|p{2cm}|}\hline
   Airline alliance  & Configuration value (p.a.a.) & Owen value (p.a.a.) & Shapley value (p.a.a.) & AENA (p.a.a.)\\\hline\hline

OneWorld&28197.656&7507.046&19825.676&20560.768\\
Ryanair&4404.690&9149.220&19083.489&18205.950\\
easyJet&2762.518&7507.028&1686.881&1742.180\\
JetStream&290.192&616.236&83.892&94.754\\
VistaJet&534.352&1245.090&147.184&163.818\\
Lufthansa&2762.518&7507.026&683.972&702.206\\
Lauda Europe&2711.038&7378.328&668.606&693.316\\
Unicair&264.224&557.806&76.646&86.682\\
NetJets Europe&383.582&843.038&54.447&61.083\\\hline		\end{tabular}}\caption{Comparison of total fees per airline alliance (p.a.a.).}\label{aggregateairlines}\end{center}
\end{table}

\section{Conclusions and further research}\label{sec:conclusions}

Realistic airport fees are key to balancing operating costs and fostering partnerships with airlines, aiming to benefit airlines and travelers alike. In this paper, we have proposed a model and a system for allocating the costs associated with the landing and the take-off of aircraft in code-sharing scenarios. To do this, we generalize the existing proposals of \cite{littlechild1973simple} and \cite{vazquez1997owen} based on airport games, now   the configuration value for games with coalition configuration \citep{albizuri2006configuration}. This study has been conducted from two different perspectives. One is purely computational: given the high computational complexity associated with this type of solutions, we have been able to provide an expression for the configuration value for airport games with coalition configuration that allows its computation in polynomial time, based on the parameters describing the underlying airport problem. To achieve this, from an axiomatic perspective, we previously consider a new characterization of the configuration value for general games with coalition configuration, based on a collection of properties that are natural in the case of airport games, among others. 

It is worth mentioning the application of this proposal to a real airport case, for which the complete list of movements of planes was available. It corresponds to the real-world situation with large set of movements of planes operating at the airport  Santiago de Compostela - Rosal\'ia de Castro  (Galicia, Spain) on the 23rd and 24th October 2024. For each of them, we have information on the set of airlines that operate it. In this setting, the configuration value was considered as a mechanism of determining a new system of aircraft fees. Besides, we compare such allocation fees with the ones obtained under two other well-known rules in airport games literature, such as the Shapley value and the Owen value, \textcolor{black}{and also with those applied by the Spanish regulatory agency for distributing the operating costs.}

Finally, we list several open issues that need to be addressed in the future from two different perspectives. One notable issue is the extension of other well-known solution concepts for TU games to the context of games with coalition configurations. Just as our proposal extends the Shapley value, with the Owen value being a special case, it would be worthwhile to develop a corresponding extension of the $\tau$-value for games with coalition configurations. This extension would also build on the proposal in \cite{casas2003extension}, which considers the presence of an a priori union system among players. Moreover, there are numerous applications of games with coalition configurations beyond the transport sector, where accounting for coalition configurations among players is both justifiable and realistic. Therefore, although we present it here in the context of determining  fees at an airport, it is important to emphasize the broader applicability of the resulting configuration value to other cost allocation problems.

\section*{Acknowledgements}

This work has been supported under grants PID2021-124030NB-C32 and PID2021-124030NB-C33, funded by MCIN/AEI/10.13039/501100011033/ and by ``ERDF A way of making Europe''/EU, and under grants \emph{Grupos de Referencia Competitiva} ED431C-2021/24  and ED431C 2020/03, funded by \emph{Conseller\'ia de Cultura, Educaci\'on e Universidades, Xunta de Galicia}.

\subsection*{Declaration of interest}
The authors declare that there is no conflict of interest.

\renewcommand{\bibfont}{\small}
\bibliographystyle{apalike}
\bibliography{referencias}

\clearpage
\appendix

\setcounter{table}{0}
\renewcommand{\thetable}{\Alph{section}.\arabic{table}}
\begin{landscape}
\section{Online Resource Section}\label{appA}
\begin{table}[!h]
	\begin{center}
		\resizebox{1.01\textwidth}{!}{	
		\begin{tabular}{|p{1.35cm}|p{1.35cm}|p{1.35cm}|p{1.35cm}|p{1.35cm}|p{1.35cm}|p{1.35cm}|p{1.25cm}|p{1.25cm}|p{1.35cm}|p{1.35cm}|p{1.35cm}|p{1.35cm}|p{1.35cm}|p{1.35cm}|p{1.25cm}|p{1.25cm}|p{1.35cm}|p{1.2cm}||p{1.5cm}||p{1cm}|p{2cm}|}	\hline
  \multirow{2}{*}{$A_1$} & \multirow{2}{*}{$A_2$} & \multirow{2}{*}{$A_3$} & \multirow{2}{*}{$A_4$} & \multirow{2}{*}{$A_5$}  & \multirow{2}{*}{$A_6$} & \multirow{2}{*}{$A_7$} & \multirow{2}{*}{$A_8$} & \multirow{2}{*}{$A_9$} & \multirow{2}{*}{$A_{10}$} & \multirow{2}{*}{$A_{11}$} & \multirow{2}{*}{$A_{12}$} & \multirow{2}{*}{$A_{13}$} & \multirow{2}{*}{$A_{14}$} & \multirow{2}{*}{$A_{15}$} & \multirow{2}{*}{$A_{16}$} & \multirow{2}{*}{$A_{17}$} & \multirow{2}{*}{$A_{18}$} & \multirow{2}{*}{$A_{19}$} & \multirow{2}{*}{CV} & \multirow{2}{*}{Type} &  {Number of movements}\\\hline\hline
54.318& & & & & & & & & & & & & & & & & &  & 54.318 &8 & 30\\
132.150 & & & & & & & & & & & & & & & & & &  & 132.150 &9 & 21\\\hdashline
  & 48.858& 57.682& & & & & & & & & & & & & & & & & 106.540 & 6 & 21\\
  & 54.006& 62.830& & & & & & & & & & & & & & & & &  116.836 & 7 & 1\\\hdashline
  & & &542.208& & & & & & & & & & & & & & &  &  542.208 & 6 &3\\
  & & & 567.947& & & & & & & & & & & & & & &  & 567.947 & 7 & 2\\\hdashline
  & 48.858& 57.682& & 129.097& & & & & & & & & & & & & &  & 235.637 & 6 &9\\
  & 54.006& 62.830& &137.677 & & & & & & & & & & & & & &  & 254.513 & 7 &2\\\hdashline
  & 54.006& 62.830& & 137.677& 559.367& 404.451 & & & & & & & & & & & &  & 1218.331 & 7 &2\\
  &48.858 & 57.682& &129.097 &542.208 & 387.291& & & & & & & & & & & &  & 1165.136 & 6&1\\\hdashline
  &48.858& & & & & & & & & & & & & & & & &  & 48.858 & 6 &5\\
  &12.115 & & & & & & & & & & & & & & & & &  & 12.115 & 5 &1\\\hdashline
  & & & & & & & 145.096 & & & & & & & & & & &  & 145.096 & 2&2\\\hdashline
  & & & & & & & & 267.176& & & & & & & & & & &267.176 & 4 &2\\\hdashline
  & & & & & & & & & 1381.259& & & & & & & & &  &1381.259 & 7&2\\\hdashline
  & 48.858& 57.682 & & & & & & & & 677.760& & & & & & & & & 784.300 & 6&2\\
  &54.006  &62.830  & & & & & & & &703.499& & & & & & & &  &  820.335& 7&2\\\hdashline
  & & & & & & & & & & & 1355.519& & & & & & &  &1355.519 &6&2\\\hdashline
  & 48.858&57.682& & 129.097& & 387.291& & & & & & 1355.519& 903.680 & 2711.038& & & &  & 5593.1668&6&1\\\hdashline
  & 48.858 & & &129.097& & & & & & & & & & & & & &  & 177.955&6&3\\
  & 54.006& & &137.677 & & & & & & & & & & & & & &  & 191.683&7&2\\\hdashline
  & 12.115 & & & & & & & & & & & & & & 121.153 & & &  &  133.268&5&5\\\hdashline
  & & & & & & & & & & & & & & & &132.112& &  &  132.112 &1&2\\\hdashline
  & & 57.682& & & & & & & & & & & & & & & &  & 57.682& 6&1\\
  & &62.830 & & & & & & & & & & & & & & & &  & 62.830&7&2\\\hdashline
  & 54.006 & 62.830& & & 559.367& 404.451& & & & & & &  955.159 & & & & &  & 2035.813&7&1\\\hdashline
  & 48.858& 57.682& & & & 387.291& & & & & & & & & & & &  & 493.831&6&1\\\hdashline
& 48.858& 57.682& &129.097& 542.208& 387.291& & & & & & 1355.519& 903.680& & & &  2711.038  & & 6135.373&6&1\\\hdashline
  & & & & & & & & & & & & & & & & & & 383.582& 383.582&3&1\\\hdashline
& & & & & & & & & & & & & & & 121.153& & & &  121.153& 5&1\\\hline\hline
4404.690	& 2762.502&	2762.534&	2762.518 &	2762.517&	2762.517&	2762.517&	290.192	&534.352	&2762.518&	2762.518	&2711.038&	2711.038&	2762.519&	2711.038&	726.918&	264.224&	2711.038&	383.582
& Total Fee& &\\\hline
\end{tabular}}
		\caption{Distribution of the fees per movement, per type of airplane and per airlines that operate it,specified by the configuration value. In last row, airline fees using Expression (\ref{CVa}).} \label{dataSCQ_costs}
	\end{center}
\end{table}
\end{landscape}

\end{document}